\documentclass[lettersize,journal]{IEEEtran}
\usepackage{amsmath,amsfonts}
\usepackage{algorithmic}
\usepackage{algorithm}
\usepackage{array}
\usepackage[caption=false,font=normalsize,labelfont=sf,textfont=sf]{subfig}
\usepackage{textcomp}
\usepackage{stfloats}
\usepackage{mathrsfs}
\usepackage{url}
\usepackage{verbatim}
\usepackage{graphicx}
\usepackage{cite}
\usepackage{multirow} 
\usepackage{makecell}
\usepackage{cellspace}
\usepackage{xcolor}
\usepackage{pifont}

\begin{document}

\title{Digital Twin Channel-Aided CSI Prediction: An Environment-Based Subspace Extraction Approach for Achieving Low Overhead and High Robustness}
\author{Yichen Cai, 
        Jianhua Zhang,~\IEEEmembership{Senior Member,~IEEE}, 
        Li Yu,~\IEEEmembership{Member,~IEEE},
        Zhen Zhang,~\IEEEmembership{Member,~IEEE},
        Yuxiang Zhang,~\IEEEmembership{Member,~IEEE},
        Lianzheng Shi,
        Yuelong Qiu, 
        Yong Zeng,~\IEEEmembership{Fellow,~IEEE}
\thanks{This work was supported by the National Natural Science Foundation of China (62525101 and 62401084), the National Key Research and Development Program of China (2023YFB2904805), and Beijing University of Posts and Telecommunications–China Mobile Communications Group Joint Innovation Center.
(Corresponding author: Jianhua Zhang.) }
\thanks{Y. Cai, J. Zhang, L. Yu, Y. Zhang, L. Shi, and Y. Qiu are with the State Key Lab of Networking and Switching Technology, Beijing University of Posts and Telecommunications, Beijing 100876, China (e-mail: \{caiyichen, jhzhang, li.yu, zhangyx, shilianzheng, yl\_qiu\}@bupt.edu.cn).}
\thanks{Z. Zhang is with the Inner Mongolia Key Laboratory of Intelligent Communication and Sensing and Signal Processing, Inner Mongolia University, Hohhot 010021, China (e-mail: zhenzhang@imu.edu.cn).}%
\thanks{Y. Zeng is with the National Mobile Communications Research Laboratory, Southeast University, Nanjing 210096, China, and also with the Purple Mountain Laboratories, Nanjing 211111, China (e-mail: yong\_zeng@seu.edu.cn).}
}


\IEEEaftertitletext{\vspace{-2.5\baselineskip}} 
\maketitle

\begin{abstract}
To meet the robust and high-speed communication requirements of the sixth-generation (6G) mobile communication system in complex scenarios, sensing- and artificial intelligence (AI)-based digital twin channel (DTC) techniques become a promising approach to reduce system overhead. In this paper, we propose an environment-specific channel subspace basis (ECB)-aided partial-to-whole channel state information (CSI) prediction method (ECB-P2WCP) for realizing DTC-enabled low-overhead channel prediction. Specifically, we introduce a wireless environment knowledge (WEK) construction method that extracts ECB from the digital twin environment via subspace estimation. This ECB characterizes the static statistical properties of the electromagnetic environment and serves as environment information prior to the prediction task. Then, we fuse ECB with real-time estimated local CSI to predict the entire spatial-frequency domain channel for both the present and future time instances. Hence, an ECB-based partial-to-whole CSI prediction network (ECB-P2WNet) is designed to achieve a robust channel prediction scheme in various complex scenarios. Simulation results indicate that incorporating ECB provides significant benefits under low signal-to-noise ratio and pilot ratio conditions, achieving a reduction of up to 50\% in pilot overhead. Additionally, the proposed method maintains robustness against multi-user interference, tolerating 3-meter localization errors with only a 0.5 dB normalized mean square error increase, and predicts CSI for the next channel coherent time within 1.3 milliseconds.
\end{abstract}

\begin{IEEEkeywords}
Digital Twin Channel, Channel Prediction, Artificial Intelligence, Channel Subspace Basis, Low Pilot Overhead.
\end{IEEEkeywords}

\vspace{-2em}
\section{Introduction}
\IEEEPARstart{I}{n} current wireless communication systems, base stations (BSs) need to obtain channel state information (CSI) by various channel estimation techniques to combat dynamic channel fading, which is a prerequisite for optimizing various transmission technologies. With the evolution of the sixth generation (6G) system towards extremely large-scale multiple-input multiple-output (XL-MIMO) and ultra-broadband transmission~\cite{lgy6G,6G,lhq}, purely relying on pilot symbols for accurate CSI acquisition will result in considerable resource overhead~\cite{ est1}. At the same time, the interference caused by massive device connections and complex wireless environment presents more stringent requirements for the robustness of channel estimation. Therefore, acquiring CSI with low overhead and strong robustness has become an important research direction.

Channel prediction is regarded as an effective way to reduce channel estimation overhead \cite{ZZ_AI,zz_drl}. Current channel prediction research can be basically divided into two categories: pilot-based or environment-based channel prediction. Pilot-based methods aim to recover the entire CSI matrix from the partially estimated CSI. Early methods relied on interpolation techniques~\cite{chazhi}, while recent studies have leveraged artificial intelligence (AI)-based methods to capture inherent channel correlations and perform inter-channel mapping effectively~\cite{14,czr_jsac,twc_est}. However, the above methods typically require accurate and sufficient channel data for prediction, and the performance may degrade when the signal-to-noise ratio (SNR) is low or the pilot ratio is significantly reduced.

Environment-based channel prediction can address the limitation of insufficient channel data by incorporating environment information~\cite{perdict,jianhua_eic,dtc}. These methods focus on extracting information such as distance, velocity, and angle from dynamic multimodal sensory data, or constructing a channel knowledge map (CKM) by mapping user locations to corresponding channel parameters~\cite{ckm2}. They support applications such as hybrid beamforming, interference suppression, and CSI prediction~\cite{ckm_CSI}, by utilizing inputs such as images~\cite{gff_camera}, point clouds~\cite{lidar}, depth maps~\cite{mmff}, or global positioning system (GPS) coordinates, all without consuming precious communication resources. However, due to the complexity and variability of the channel environment, the current environment-based channel prediction is very sensitive to the user's location, and its performance is easily affected. There is a lack of CSI prediction algorithms that are applicable to various line-of-sight (LoS) and non-line-of-sight (NLoS) scenarios and robust to localization errors.

In addition, advances in 6G sensing technology enable communication systems to be equipped with GPS, cameras, and light detection and ranging (LiDAR), making it possible to achieve environmental intelligent communication powered by the digital twin~\cite {jianhua_eic}. In \cite{dtc}, the digital twin channel (DTC) can more effectively utilize rich environment information to enhance robustness. DTC predicts the physical channel by reconstructing a high-precision 3D environment from multimodal sensory data and leveraging AI to comprehensively predict channel propagation characteristics in the digital domain. Essentially, channel fading arises from the interaction of electromagnetic waves with scatterers in the physical environment, such as buildings, trees, and pedestrians. Therefore, variations in CSI are primarily determined by the surrounding physical environment, making environment information a valuable prior knowledge for CSI prediction~\cite{SYT_feature,4steps}. By leveraging environmental prior knowledge, DTC reduces the reliance on pilot signals, offering a promising approach to minimizing communication overhead. In addition, to reduce redundancy in sensory data and enhance the efficiency of channel prediction, the wireless environment knowledge (WEK) is adopted, which refers to the process of extracting interpretable, channel-relevant knowledge from the environment~\cite{REKP}. 

In DTC-based channel prediction, existing research primarily focuses on large-scale tasks such as path loss (PL) prediction~\cite{REKP}, where the information extracted through WEK primarily captures coarse-grained spatial features. Such representations, while effective for macroscopic channel characteristics, lack the fine-grained spatial–frequency features required for accurate small-scale CSI prediction. Therefore, effectively realizing the DTC paradigm for robust and low-overhead CSI prediction still faces two key challenges: what channel-relevant environment information needs to be extracted to support CSI prediction, and how to incorporate such information into AI-based prediction frameworks in a simple and efficient manner.

Motivated by this, we propose an environment-specific channel subspace basis (ECB)-aided partial-to-whole CSI prediction method (ECB-P2WCP) to achieve low-overhead and robust channel prediction. The extracted ECB from the environment exhibits strong correlation with the target CSI to be predicted. The main contributions of this paper are summarized as follows:

$\bullet$ A general framework for low-overhead CSI prediction is presented, which can efficiently utilize online environment priors with minimal additional overhead by integrating both static and dynamic sensing data within the DTC paradigm. A structured WEK applicable to arbitrary deployment scenarios is constructed, from which ECB is extracted. By leveraging pre-constructed static environments in DTC and low-density spatial sampling simulations, ECBs can capture channel statistics under both static and mobile conditions and be extracted efficiently at minimal cost. The proposed ECB construction method is not constrained by scene layout and avoids frequent environment updates. In particular, it is insensitive to localization errors.

$\bullet$ We design a robust ECB-based partial-to-whole CSI prediction network (ECB-P2WNet) that effectively integrates ECB and partial CSI to extract high-level channel representations. The network supports both present and future CSI prediction within a unified architecture. Specifically, ECB-P2WNet predicts the present CSI via a convolutional neural network (CNN)- and Transformer-based network and forecasts future CSI by appending a long short-term memory (LSTM) module. The overall architecture remains unchanged, and network parameters are shared across tasks, enabling temporal prediction without increasing model complexity.

$\bullet$ Comprehensive simulation results show that the proposed method achieves superior prediction performance as well as robustness to multi-user interference and localization errors. Under low-SNR and low-pilot-overhead conditions, the gains provided by environment information are particularly significant, saving up to 50\% of pilot overhead. Moreover, the proposed method can predict the CSI of the next coherent time within 1.3 milliseconds (ms), effectively mitigating channel aging while further reducing pilot overhead.

The remainder of this paper is organized as follows: Section \uppercase\expandafter{\romannumeral2} introduces the channel and WEK models and formulates the CSI prediction problem. Section \uppercase\expandafter{\romannumeral3} presents the details of the WEK construction and ECB-P2WNet. Section \uppercase\expandafter{\romannumeral4} shows the simulation results and Section \uppercase\expandafter{\romannumeral5} draws the conclusion.

\emph{Notations:} Lowercase letters $x$ denotes scalar, boldface lowercase letters $\mathbf{x}$ denotes column vector, and boldface uppercase letters $\mathbf{X}$ denotes matrix. The $m$-th entry of a column vector $\mathbf{x}$ and the entry at the $m$-th row and $n$-th column of a matrix $\mathbf{X}$ is denoted by $\left[\mathbf{x}\right]_m$ and $\left[\mathbf{X}\right]_{m,n}$, repectively. The vector corresponding to the $m$-th row and the $n$-th column of a matrix $\mathbf{X}$ is denoted by $\mathbf{X}_{m,:}$ and $\mathbf{X}_{:,n}$, respectively. The conjugate transpose is represented as $\left(\cdot\right)^H$. The operator $\|\cdot\|_p$ denotes the $p$-norm, while $\odot$ represents the Hadamard product.

\vspace{-1em}
\section{System Model and Problem Formulation}
\begin{figure}[t]
\centering
\includegraphics[width=0.4\textwidth]{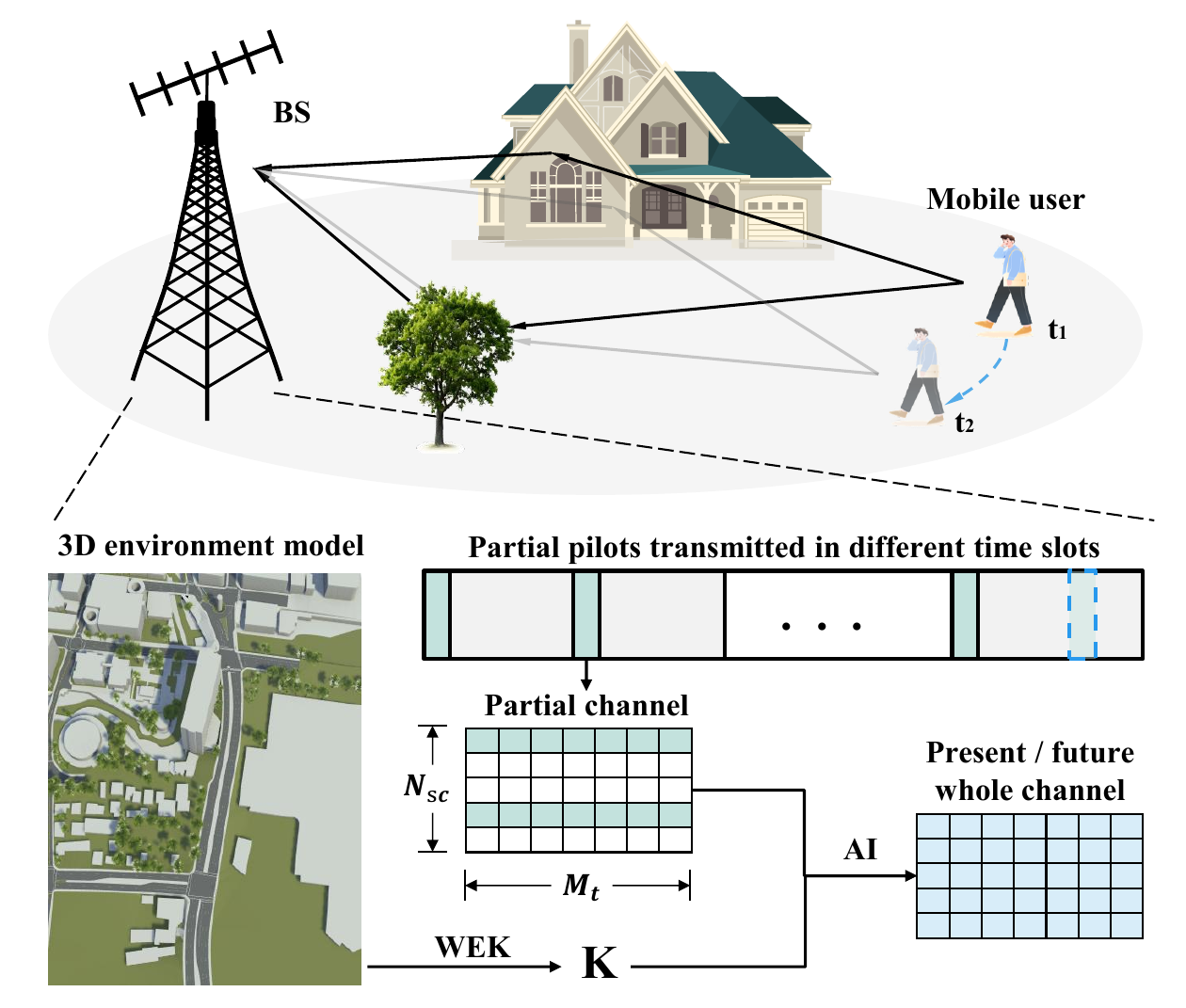}
\caption{The method of obtaining partial CSI in a MIMO-OFDM system.}
\label{sys}
\end{figure}
Let us consider a single-cell time division duplex (TDD) MIMO system employing orthogonal frequency-division multiplexing (OFDM) mode. The main objective is to predict the entire CSI from partial channel estimation results and environment information, thereby reducing the uplink pilot overhead, particularly the sounding reference signal (SRS) transmitted by the user, as illustrated in Fig.~\ref{sys}.

\vspace{-1em}
\subsection{System Model}
The BS is equipped with a uniform planar array (UPA) of $M_t$ antennas ($M_t \gg 1$) to communicate with a single-antenna user, and operates with OFDM modulation using $N_{sc}$ subcarriers. A geometric channel model is adopted~\cite{zz5}, where the channel of the $k$-th subcarrier in the $t$-th channel coherence interval is given by:

\begin{equation}
\label{deqn_ex1a3}
\mathbf{h}_{k,t}=\sum_{l=1}^{L}\sqrt{\frac{\alpha_{l,t}}{N_{sc}}}e^{j(\frac{2\pi k\tau_{l,t}B}{N_{sc}}+\psi_{l,t})}\mathbf{a}\left( \theta_{l,t}, \phi_{l,t} \right),
\end{equation}
where $\mathbf{h}_{k,t} \in \mathbb{C}^{M_t }$ denotes the channel vector. The parameters $\alpha_{l,t}$, $\tau_{l,t}$, $\psi_{l,t}$, $\theta_{l,t} \in \left[ -\frac{\pi}{2}, \frac{\pi}{2} \right]$, and $\phi_{l,t} \in [0, 2\pi)$ represent the received power, delay, phase, elevation angle, and azimuth angle of the $l$-th path, respectively. $B$ denotes the channel bandwidth, and $\mathbf{a}\left( \theta_{l,t}, \phi_{l,t} \right)$ represents the steering vector~\cite{deepmimo}. 

When the user transmits a uplink pilot symbol $x_k \in \mathbb{C}$ on the $k$-th subcarrier, the received signal $\mathbf{y} _{k,t}\in \mathbb{C}^{M_t}$ at the BS can be expressed as:
\vspace{-0.5em}
\begin{equation}
\label{deqn_ex1a4}
\mathbf{y}_{k,t}=\mathbf{h}_{k,t}x_{k,t}+\mathbf{n}_{k,t},\quad k=1,\dots,N_{sc},
\end{equation}
where ${\mathbf{n}_{k,t}} \in {\mathbb{C}^{{{{M}}_{{t}}}}}$ is the zero-mean circularly symmetric complex Gaussian (CSCG) noise with variance $\sigma^2$.

By stacking all subcarriers, the whole channel response matrix ${\mathbf{H}_t}\in {\mathbb{C}^{{{{M}}_{{t}}} \times {N_{sc}}}}$ in the spatial and frequency domain can be expressed as:
\vspace{-0.5em}
\begin{equation}
\label{deqn_ex1a5}
{\mathbf{H}_t}=[\mathbf{h}_{1,t},\mathbf{h}_{2,t},...,\mathbf{h}_{N_{sc},t}].
\end{equation}

To reduce pilot overhead and enhance the data transmission rate, the pilot signal transmitted by the user is selectively placed on $N_p$ subcarriers ($N_p \ll N_{sc}$), instead of occupying all subcarriers. In this case, the partial channel matrix $\mathbf{H}^{0}_t\in \mathbb{C}^{{M_t}\times N_{p}}$, obtained from the $N_p$ pilot positions, can be regarded as a incomplete observed version of $\mathbf{H}_t$, denoted as:
\vspace{-0.5em}
\begin{equation}
\label{deqn_ex1a6}
\mathbf{H}^0_t=\mathbf{B} \odot \mathbf{H}_t,
\end{equation}
where $\mathbf{B} \in \{0,1\}^{M_t \times N_{sc}}$ is a binary observation matrix, and $\|\mathbf{B}_{i,:}\|_0 = N_p$, $\forall i \in \{0, \dots, M_t-1\}$. Under the considered system model, all $\mathbf{B}_{i,:}$ are identical.



\subsection{Wireless Environment Knowledge Model}
To establish a task-oriented connection between environmental data and channel characteristics, the concept of WEK within the DTC framework is introduced. WEK denotes the interpretable relationship mapping $K(\:\cdot\:)$ that treats the environment information ${\mathbf E} \in {\mathbb{R}^{{D}}}$ to the channel knowledge matrix ${\mathbf K} \in {\mathbb{C}^{J_1 \times J_2}}$, expressed as:
\begin{equation}
\label{deqn_ex1a7}
\mathbf{K}=\begin{Bmatrix}K\left(\:\cdot\:\right):\mathbb{R}^{D}\to\mathbb{C}^{J_1 \times J_2}\end{Bmatrix},
\end{equation}
where $\mathbf{E}$ represents $D$-dimensional environment information obtained from the physical world, such as the position, size, and spatial layout of surrounding scatterers. The output $\mathbf{K}$ encodes interpretable, channel-related knowledge matrix, where the dimensionality $J_1$ and $J_2$ is determined by the requirements of the target task, such as angular resolution, delay spread representation, or frequency granularity~\cite{REKP}.

Once constructed, the WEK remains unchanged throughout system deployment. However, its output $\mathbf{K}$ dynamically adapts to the varying environment input $\mathbf{E}$.

\subsection{Problem Formulation}
Then, the low-overhead channel prediction task is formulated. Based on the WEK model, the objective is to design a task-oriented WEK construction method that generates an appropriate knowledge matrix $\mathbf{K}$. Given $\mathbf{K}$ and the observed $\mathbf{H}^0$, a prediction algorithm is developed to estimate $\mathbf{H}$. The present whole CSI matrix $\mathbf{\hat H}_t\in \mathbb{C}^{{M_t}\times N_{sc}}$ prediction problem can be formulated as:

\begin{equation}
\label{e1}
Problem\text{ }1:\text{ }(\mathbf{H}^0_t,\mathbf{K}_t)\stackrel{M(\cdot)}{\longrightarrow} \mathbf{\hat{H}}_t,
\end{equation}
where $M(\cdot)$ denotes the frequency prediction function.

To address the issue of channel aging, a temporal prediction function $G(\cdot)$ is introduced to predict the future CSI matrix $\mathbf{H}_{t+1}$ based on the results of previous $T$ time instances, formulated as:

\begin{equation}
\label{eg}
Problem\text{ }2:\text{ }(\mathbf{H}^0_{t-T+1}, \dots, \mathbf{H}^0_t, \mathbf{K}_t)\stackrel{G(\cdot)}{\longrightarrow} \mathbf{\hat{H}}_{t+1}.
\end{equation}

To avoid error accumulation during recursive prediction, the estimated $\mathbf{\hat{H}}_t$ from Eq.~\eqref{e1} is not used as input to the temporal predictor. Instead, the actual observed $\mathbf{H}^0_t$ is employed to ensure stability.

\section{THE PROPOSED ECB-P2WCP}
In this section, a detailed description of ECB-P2WCP is provided. First, the workflow is designed based on specific task requirements. Subsequently, the WEK is constructed through an analysis of ray-tracing (RT) data to realize the extraction of ECB. Building on this foundation, a deep learning (DL) model is developed to achieve low-overhead CSI prediction, followed by a detailed explanation of ECB-P2WNet’s procedure and its constituent components.

\subsection{Motivation and Architecture}
As a starting point, we focus on the key challenge of effectively decomposing diverse and large-scale environment data to extract $\mathbf K$ that are most informative for the CSI prediction task. Several considerations arise in this process.

\textbullet\ Firstly, to enhance the utility of applying environment information, the extracted $\mathbf{K}$ should be directly relevant to the channel characteristics, rather than representing only geometric details such as positions, sizes, or materials of buildings, which lack a direct and reliable mapping to CSI.

\textbullet\ Secondly, to avoid the massive and impractical engineering effort of extracting a distinct $\mathbf{K}$ for every position within a cell, the scene is uniformly divided into grids with several meters side lengths, reflecting the distinct physical environmental characteristics of different local subregions~\cite{scene-specific}. Accordingly, all users within the same grid should be associated with a common $\mathbf{K}$.

\textbullet\ Moreover, unlike large-scale channel parameters, such as PL, which change relatively slowly over space, small-scale parameters, such as phase, can undergo significant changes within a few meters. To make $\mathbf K$ effective for all positions within a grid, spatial stationarity of $\mathbf K$ is desired. 

\textbullet\ Additionally, the extracted $\mathbf K$ is expected to remain time-invariant. This ensures that during actual deployment, the system does not require frequent real-time updates, thereby avoiding additional resource consumption.

\begin{figure*}[t]
\centering
\includegraphics[width=0.9\textwidth]{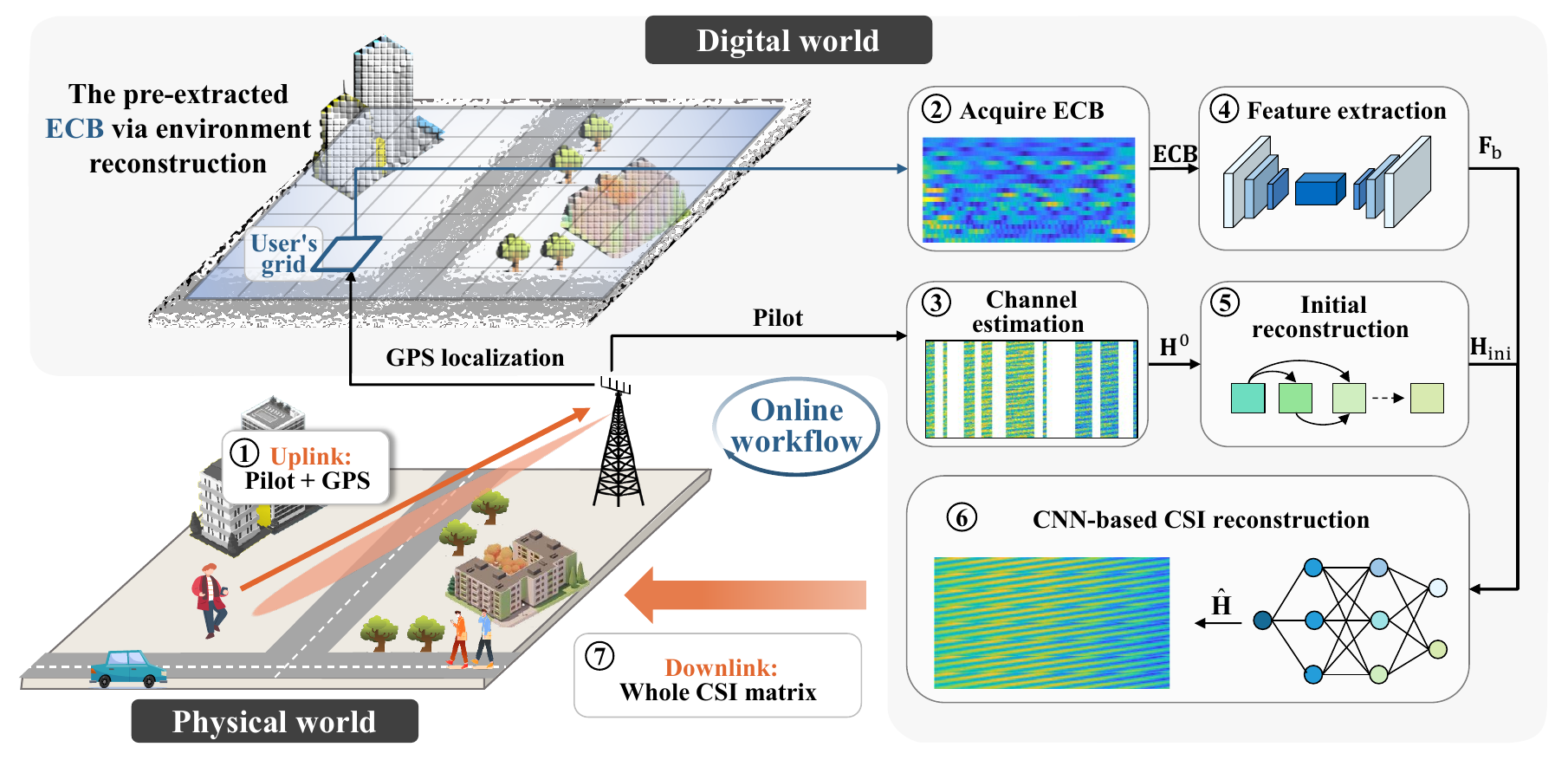}
\vspace{-1em}
\caption{The workflow of the low-overhead CSI prediction process combining static and dynamic data.}
\label{arc}
\end{figure*}

To meet these requirements, $\mathbf{K}$ should reflect the static statistical characteristics of different regions, allowing it to generalize to users at various locations. At the same time, the second key challenge lies in how to utilize $\mathbf K$ for AI-based channel prediction efficiently. The model should be able to dynamically select the region-specific $\mathbf K$, while incorporating online information to account for environmental variations. Prior studies have demonstrated that the channel subspace is strongly tied to the underlying physical environment ~\cite{finger}, which supports the design choice of extracting the basis vectors of channel subspace for each grid as $\mathbf K$, referred to as ECB. Building on this, we propose a DTC framework that combines both static and dynamic data to enable low-overhead CSI prediction. As the user moves within the environment, the static component is realized by reconstructing the scene preliminarily and extracting ECB as a compact and high-level abstraction of environmental information. Although $\mathbf{H}$ changes over time and space, it can always be represented by a linear combination of the basis vectors. The dynamic component leverages real-time information, such as GPS coordinates, to track user movement and partial uplink pilot signals, reflecting instantaneous channel variation. 

\begin{figure}[t]
\centering
\includegraphics[width=0.4\textwidth]{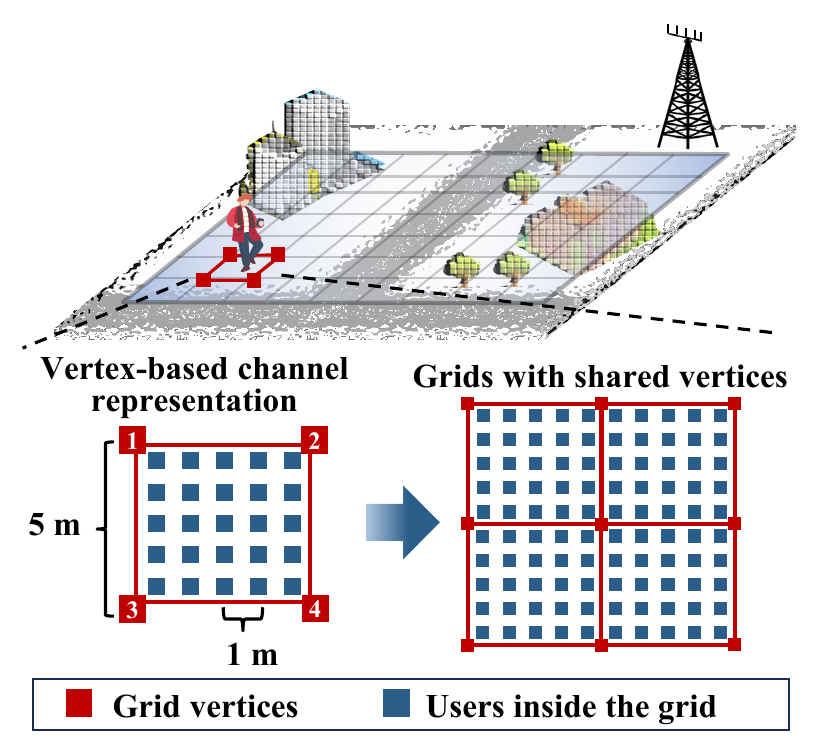}
\vspace{-1em}
\caption{The proposed ECB extraction method.}
\label{weks}
\end{figure}

The complete workflow of the proposed ECB-P2WCP is illustrated in Fig.~\ref{arc}. In each cycle, the user transmits pilot signals along with their current GPS coordinates to the BS (step 1). The GPS information is used to identify the corresponding ECB in the digital world (step 2), while the pilot is processed to obtain the partial $\mathbf{H}^0$ via channel estimation (step 3). These two inputs are then jointly fed into the DL model to predict the complete $\mathbf{\hat{H}}$ (steps 4–6). Finally, the predicted $\mathbf{\hat{H}}$ is delivered to the physical world (step 7), enabling its use in air interface procedures such as downlink precoding. This framework enables the exploitation of environment-specific information with minimal overhead, without relying on additional sensing or communication resources.

\vspace{-1em}
\subsection{WEK Construction and ECB Extraction}\label{eb}
In this subsection, the structural WEK is constructed to enable ECB extraction. By flattening the spatial–frequency domain channel matrix $\mathbf{H}$ into a vector $\mathbf{h}_{\text{flat}} \in \mathbb{C}^{N_H}$, with $N_H = M_t \times N_{sc}$, the channel can be represented in an $N_H$-dimensional space. However, storing all $N_H$ orthogonal basis vectors for the entire channel subspace is unnecessary and inefficient. In practice, under limited scattering conditions, the channel often exhibits a low-rank structure~\cite{gff_rank}. This implies that $\mathbf{h}_{\text{flat}}$ can be effectively approximated using only $N_b$ dominant basis vectors corresponding to the largest projection coefficients~\cite{zjj}, expressed as:

\begin{equation}
\label{deqn_ex1a8}
\mathbf {h}_{\text{flat}}=\mathbf{K}\mathbf {c},
\end{equation}
where $\mathbf{K}\in {\mathbb{C}^{N_H \times N_{b}}}$ represents a set of orthogonal basis in the channel subspace and $\mathbf {c}\in {\mathbb{C}^{N_{b}}}$ denotes the corresponding coefficients.

\begin{figure*}[t]
\centering
\includegraphics[width=0.95\textwidth]{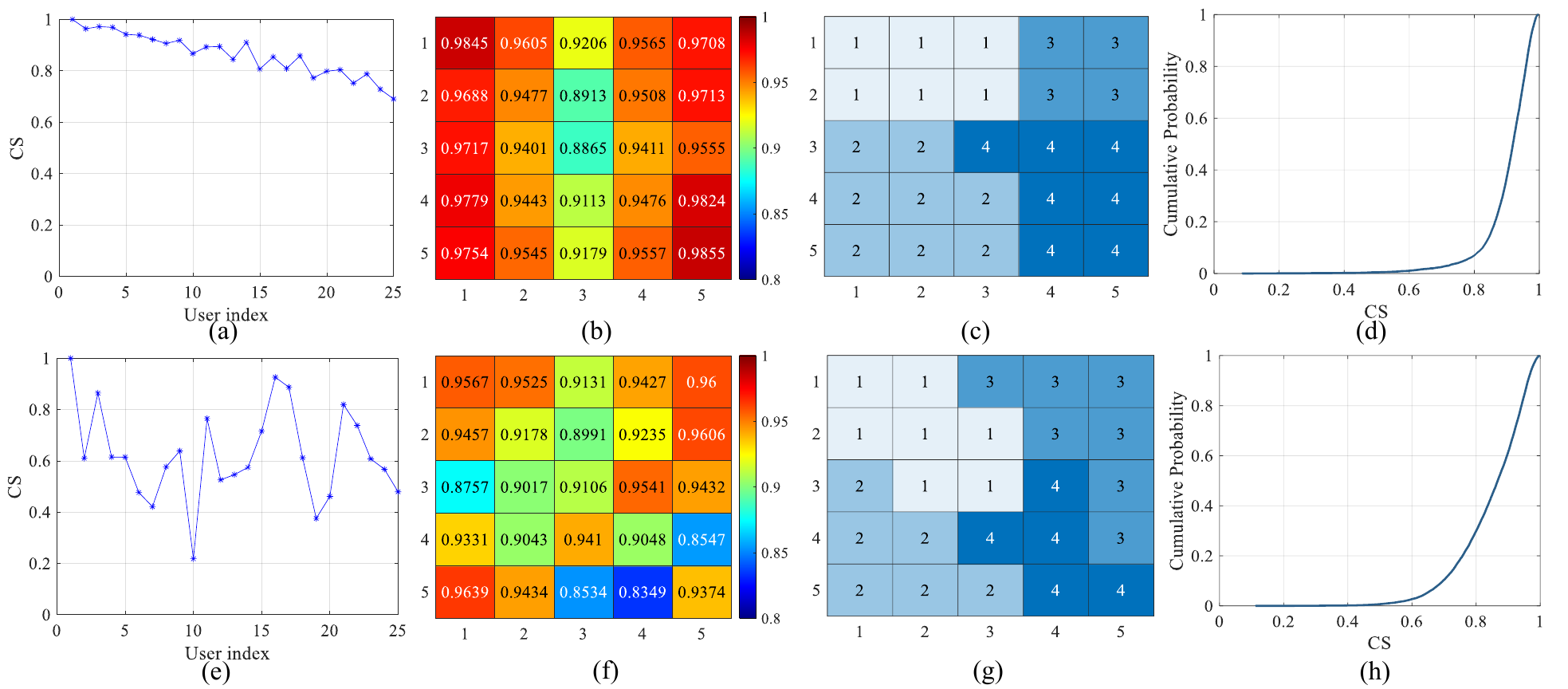}
\vspace{-1em}
\caption{Correlation between channels at different locations. a) CS between channel in the LoS grid; b) Maximum CS between user channel and vertex channel in a LoS grid; c) Index of vertex with maximum CS to user channel in a LoS grid; d) CS distribution of all the LoS position channels with vertex channels; e)-h) the same result in the NLoS region.}
\label{corrh}
\end{figure*}

Inspired by \cite{prior_csi}, a limited number of CSI samples from selected positions can represent the channel characteristics within a spatial grid. Accordingly, the $N_b$ basis vectors can be constructed from the spatial basis of these samples. 
To reduce the overall measurement or simulation overhead by enabling CSI sample sharing across adjacent grids, ECB for each grid is constructed using the channels at its four vertex points (highlighted in red in Fig.~\ref{weks}). To capture the impact of user mobility on the channel subspace, Doppler shifts corresponding to a pedestrian speed of 3 kilometers per hour (km/h) are incorporated. Channel samples are taken at both short- and long-term time instances, specifically at $t = \left\{ 0,\ 40,\ 80,\ \dots,\ 360 \right\}~\text{ms}$. Consequently, each grid obtains $4 \times 10$ space-frequency domain channel realizations from its four vertices. 

To validate the representativeness of the vertex-based channels, an outdoor urban macrocell (UMa) simulation scenario is set up to emulate the environment reconstruction process in DTC, and RT simulations are performed to generate the corresponding channel data. In the three-dimensional (3D) environment model, only static objects are considered. Although pedestrians and vehicles introduce some dynamism in real deployments, their impact on the channel is significantly smaller than that of buildings and roads~\cite{zy3}, resulting in a negligible effect on the extraction of the ECB. Since~\cite{lqs} showed that local quasi-stationarity (LQS) regions in UMa scenarios typically extend over several meters up to more than ten meters, and~\cite{ckm_qyl} verified that meter-level grids preserve channel statistics, the grid side length is set to 5 meters (m). Within each grid, 25 users (marked in blue in Fig.~\ref{weks}) are uniformly distributed within each grid to simulate users in diverse spatial locations, and their corresponding space-frequency domain channels are also generated. To facilitate the analysis, one grid is selected from each of the LoS and non-line-of-sight (NLoS) regions. Specifically, Figs.~\ref{corrh}(a)–(d) present the results for the LoS grid, while Figs.~\ref{corrh}(e)–(h) correspond to the NLoS grid. The channel correlation between different users is evaluated, focusing on the relationship between the user located at the upper-left corner and the remaining 24 users. Cosine similarity (CS) is adopted as the correlation metric, defined as:

\begin{equation}
\label{eq:cos_sim}
\text{CS} = \frac{|\mathbf{h}_{\text{flat},1}^H \mathbf{h}_{\text{flat},2}|}{\|\mathbf{h}_{\text{flat},1}\|_2 \|\mathbf{h}_{\text{flat},2}\|_2},
\end{equation}
where $\mathbf{h}_{\text{flat},1}$ and $\mathbf{h}_{\text{flat},2}$ represent the two channels under correlation analysis. As shown in Figs.~\ref{corrh} (a) and (c), the CS in the LoS region generally decreases as distance increases. In contrast, no such pattern is observed in the NLoS region, where the CS across different positions is generally lower.

Next, each user channel is compared against the 40 vertex channels within the grid, and the maximum CS along with the corresponding vertex index is recorded, where the vertex indexing follows the convention shown in Fig.~\ref{weks}. As illustrated in Figs.~\ref{corrh}(b), (c), (f), and (g), although user channels vary across locations, there consistently exists at least one vertex channel that exhibits high similarity with the corresponding user channel. The cumulative distribution function (CDF) of the maximum CS across all users in both the LoS and NLoS regions is presented in Figs.~\ref{corrh}(d) and (h), respectively. At the 50\% CDF point, the CS reaches 0.92 in the LoS region and 0.87 in the NLoS region. The results demonstrate that the vertex channels are capable of representing the dominant channel behavior within each grid, validating the feasibility of ECB extraction from such observations.

Then, a subspace estimation method based on eigenvalue decomposition (EVD) is employed to identify the most significant basis vectors~\cite{pca}. The $N_H \times N_H$ autocorrelation matrix of the vertex channels is calculated and then decomposed by EVD, expressed as :

\begin{equation}
\label{deqn_ex1a9}
EVD\left( {\sum\limits_{i = 1}^4 {\sum\limits_{j = 1}^{{t_m}} {\mathbf {h}_ {_{\text{flat}}}^{i,j}{{\left( {\mathbf {h}_ {_{\text{flat}}}^{i,j}} \right)}^H}} } } \right)={\mathbf{U\Lambda U }}^H,
\end{equation}
where $\mathbf {{h}}_{\text{flat}}^{i,j}$ represents the channel at the $i$-th vertex at the $j$-th time, and $t_m = 10$ is the maximum time considered. Besides, $\mathbf {U}$ and ${\Lambda}\in {\mathbb{C}^{N_H \times N_H}}$ denote the eigenvector matrix and the diagonal matrix containing the eigenvalues, respectively.

Then, the number of retained basis $N_{b}$ is selected based on the cumulative energy ratio of the first $\alpha$ eigenvalues, calculated as:

\begin{equation}
\label{deqn_ex1a10}
\eta_{\alpha}=\frac{\sum_{i=1}^{\alpha}\lambda_i}{\sum_{i=1}^{N_H}\lambda_i},
\end{equation}
where $\lambda_i$ represents the $i$-th eigenvalue arranged in descending order.

\begin{figure}[t]
\centering
\includegraphics[width=0.45\textwidth]{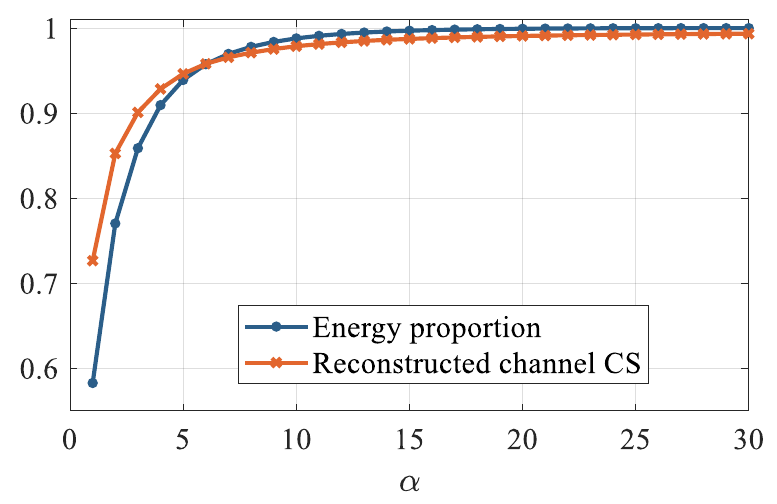}
\vspace{-1em}
\caption{Energy proportion and reconstructed channel CS.}
\label{bianli}
\end{figure}

By traversing each grid overall region, the average $\eta_{\alpha}$ is computed, as shown by the blue line in Fig.~\ref{bianli}. It is evident that the first fifteen eigenvalues account for about 99\% of the total energy. Besides, the performance of reconstructing $\mathbf {h}_{\text{flat}}$ using the first ${\alpha}$ basis is also evaluated to validate the energy-based selection of $N_{b}$. First, the projection of $\mathbf {h}_{\text{flat}}$ in the chosen basis space is computed as:

\vspace{-0.5em}
\begin{equation}
\label{e2}
\mathbf {\hat{\mathbf {c}}} = \mathbf {h}_{\text{flat}}{\mathbf {U}_\alpha},
\end{equation}
where $\mathbf {U}_{\alpha}\in {\mathbb{C}^{N_H \times {\alpha}}}$ is the matrix containing the top-${\alpha}$ eigenvectors, $\hat{\mathbf {c}}\in {\mathbb{C}^{{\alpha}}}$ represents the projection coefficients. Then, reconstruct the original signal sparsely after projection, obtaining the reconstructed data $\hat {\mathbf {h}}_\text{flat}$:

\vspace{-0.5em}
\begin{equation}
\label{e3}
\hat {\mathbf {h}}_{\text{flat}} = \hat{\mathbf {c}}{\mathbf {U}_\alpha^H}.
\end{equation}

The average reconstruction CS for varying values of ${\alpha}$ is shown by the red line in Fig.~\ref{bianli}. It closely follows the trend of the eigenvalue energy distribution. Since the CS shows negligible improvement beyond ${\alpha} = 15$, the number of basis vectors is set to $N_b = 15$. Accordingly, the WEK is constructed as $\mathbf{K} = \mathbf{U}_{15}$. For each grid with distinct environment characteristics, a corresponding ECB is pre-extracted and stored in the digital world. The ECB captures structured channel patterns specific to the environment and serves as a useful prior for online CSI prediction.

\vspace{-1em}
\subsection{ECB-Aided CSI Prediction Network Design}
A simple approach to implementing the ECB-aided CSI prediction process is to employ a projection-based reconstruction algorithm. Specifically, the input to Eq.~\eqref{e2} is replaced with the flattened partial channel matrix $\mathbf {h}^0_{\text{flat}} \in {\mathbb{C}^{N_H}}$, and Eq.~\eqref{e3} is subsequently used to obtain $\mathbf{\hat{H}}$. As demonstrated in Section~\ref{eb}, this method yields relatively high CS when $\mathbf{H}$ and $\mathbf U$ are noiseless and fully sampled. However, in practical scenarios, both $\mathbf U$ and $\mathbf H$ are affected by noise, which can lead to performance degradation when projected into the signal subspace. To address this, a DL-based network, ECB-P2WNet, is designed to implement the frequency mapping function ${M(\cdot)}$ in Eq.~\eqref{e1}, and further realize the temporal prediction function ${G(\cdot)}$ defined in Eq.~\eqref{eg}. 

\begin{figure*}[t]
\centering
\includegraphics[width=0.85\textwidth]{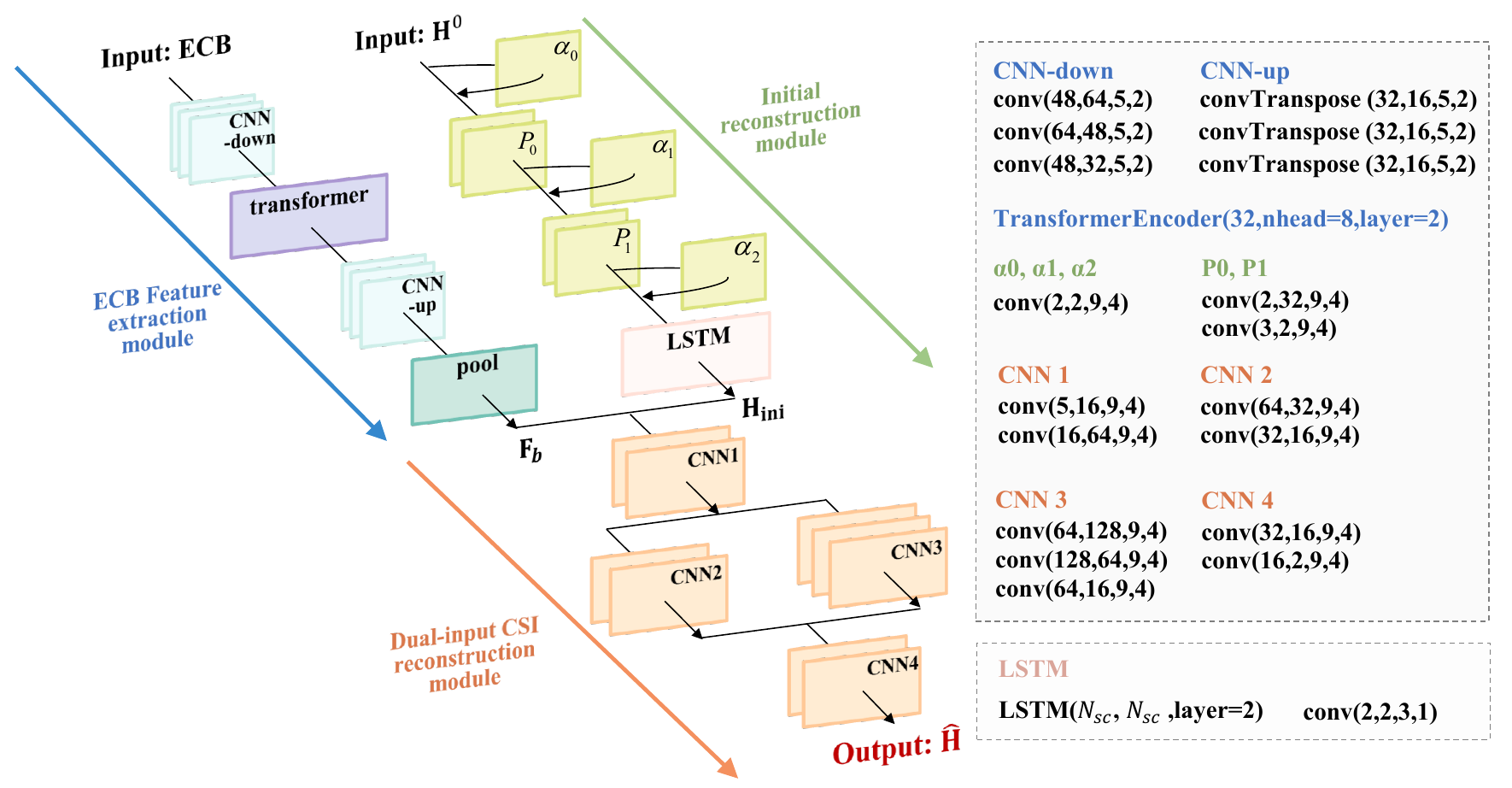}
\vspace{-1em}
\caption{The architecture of the proposed ECB-P2WNet}
\label{net}
\end{figure*}

As shown in Fig.~\ref{arc}, the ECB-P2WNet consists of three subnetworks: ECB feature extraction subnetwork (step 4), initial reconstruction subnetwork (step 5), and dual-input CSI reconstruction subnetwork (step 6). Their architectures and parameters are detailed in Fig.~\ref{net}, where the LSTM~\cite{lstm} module is employed only for the future CSI prediction task.

\textit{1) ECB feature extraction subnetwork: }This subnetwork is designed to transform ECB into a latent representation that is aligned with the latent space of the network. The output is the ECB feature matrix $\mathbf{F}_b$, which has the same dimensions as $\mathbf{H}^0$, facilitating subsequent feature concatenation.

The network begins with several convolutional layers ~\cite{cnn} to capture local spatial–frequency patterns. Dimensionality reduction is achieved by setting the convolution stride to 2, which reduce data redundancy and complexity, thereby improving the efficiency of subsequent processing. Following this, a transformer encoder~\cite{transformer} is employed to capture long-range dependencies and complex spatial–frequency interactions among features. This enhances the model’s ability to integrate global information and extract essential features for accurate CSI recovery.

The encoded features are then reshaped to match the spatial–frequency dimensions of $\mathbf H$. Transposed convolutional layers are used for upsampling, reconstructing higher-resolution feature maps that preserve detailed information. Furthermore, each convolutional and transposed convolutional layer is followed by a batch normalization layer, which mitigates vanishing and exploding gradients, accelerates convergence, and improves training stability and generalization.

\textit{2) Initial reconstruction subnetwork:} The input to this subnetwork is $\mathbf{H^{0}}$, and the output is the initially reconstruction result $\mathbf{H}_\text{ini}$.
Recovering $\mathbf{H}_\text{ini}$ from the partially known values $\mathbf{H^{0}}$ can be formulated as a matrix completion problem, which is solved using the proximal gradient algorithm~\cite{dps}. In each iteration, the gradient of the loss function is computed, followed by a proximal operation to obtain the updated estimate, expressed as:
\begin{equation}
\label{deqn_ex1a11}
\mathbf{H}_\text{ini}^{\left(l + 1\right)} 
= \mathcal{P}\!\left( \mathbf{H}_\text{ini}^{(l)} 
- \gamma_l \nabla f\!\left(\mathbf{H}_\text{ini}^{(l)}\right) \right),
\end{equation}
where $f({\mathbf{H}}_\text{ini}^{{(l)}})=\frac{1}{2}\| {\mathbf{B}}({{\mathbf{H}}_\text{ini}} - {{\mathbf{H}^0}}){\| ^2}$ denotes the reconstruction error, ${\cal{P}} (\cdot)$ and $\gamma_l$ represent the proximal mapping and stepsize at the $l$-th iteration, respectively. To implement this iterative algorithm, a CNN-based network is employed, where learning is conducted at each iteration. The update rule is reformulated as:
\begin{equation}
\label{deqn_ex1a12}
{\mathbf{H}}_\text{ini}^{{(l + 1)}} = {\cal P}_\theta ^{(l)}\left( {{\mathbf{H}}_\text{ini}^{{(l)}} - \gamma _\theta ^{(l)}\left( {{{\mathbf{B}}}\odot{\mathbf{H}}_\text{ini}^{{(l)}} - {{\mathbf{H}}^0}} \right)} \right),
\end{equation}
where ${\cal P}_\theta ^{(l)}$ is modeled by a two-layer CNN, and $\gamma _\theta ^{(l)}$ is modeled by a one-layer CNN. The number of iterations is set to $L=2$. In each iteration, the CNN learns to refine the approximation of $\mathbf{H_\text{ini}}$ by minimizing the reconstruction error, progressively improving the estimate of the whole channel matrix based on the available partial observations.

\textit{3) Dual-input CSI reconstruction subnetwork:} This subnetwork combines the feature matrix $\mathbf{F}_b$ with the initially reconstruction result $\mathbf{H}_\text{ini}$ to perform the final CSI reconstruction. Specifically, $\mathbf{F}_b$ and $\mathbf{H}_\text{ini}$, which are spatially aligned, are concatenated along the channel dimension and subsequently fed into a CNN-based network. 

The network employs residual blocks to facilitate information flow and preserve key features across layers~\cite{resnet}. By integrating outputs from both shallow and deep paths, the network maintains low-level details while enabling deeper-level refinement. This architecture enables the extraction of multi-level spatial–frequency features, where shallow layers capture simple fluctuations and deep layers learn complex variations, ultimately improving CSI reconstruction accuracy.

\begin{figure}[t]
\centering
\includegraphics[width=0.45\textwidth]{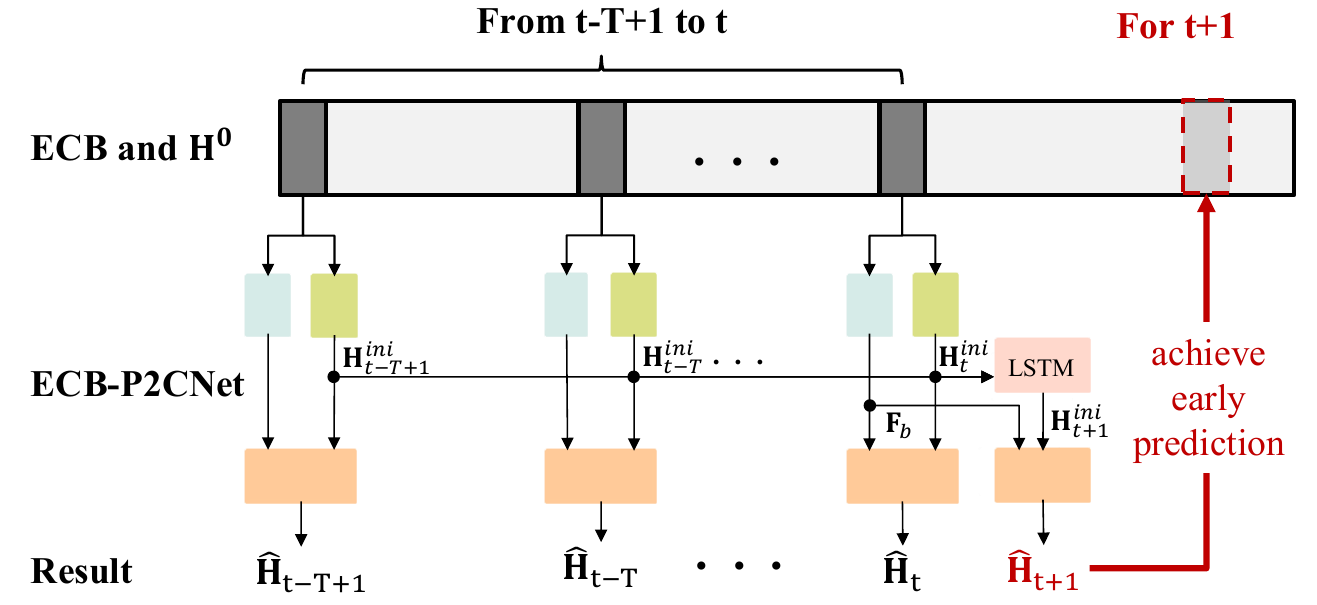}
\vspace{-1em}
\caption{ECB-P2WNet usage across time instances.}
\label{pert}
\end{figure}

\textit{4) Training process: }First, the network for the present CSI prediction task is trained by jointly optimizing the parameters of the three subnetworks. The loss function is defined as the mean squared error (MSE) between the predicted and actual CSI values, expressed as:
\begin{equation}
\label{deqn_ex1a13}
\mathcal{L}_{\text{MSE}} = \frac{1}{M_t N_{sc}} \left\| \mathbf{H} - \hat{\mathbf{H}} \right\|_F^2.
\end{equation}

Subsequently, the network is adapted to the future CSI prediction task. To minimize model storage requirements across tasks, the overall architecture is kept unchanged. Instead, an LSTM module is appended to the end of the initial reconstruction subnetwork to capture the temporal correlation of CSI, as shown in the pink region of Fig.~\ref{net}. The prediction process is illustrated in Fig.~\ref{pert}. Specifically, $\mathbf{H}_\text{ini}$ from the previous $T$ time instants is collected and fed into the LSTM to generate $\mathbf H_{\text{ini},{T+1}}$, which is then processed by the dual-input CSI reconstruction subnetwork to produce the final prediction $\hat{\mathbf{H}}_{T+1}$. During deployment, historical $\mathbf{H}_\text{ini}$ and $\mathbf{F}_{b}$ can be retained from previous predictions, avoiding redundant computation in the initial reconstruction and ECB feature extraction subnetworks. As a result, the added prediction task introduces minimal computational and storage overhead.

The training is conducted in two stages. First, the pre-trained parameters are frozen, and only the LSTM module is updated using the MSE loss between $\mathbf{H}_{t+1}$ and $\hat{\mathbf{H}}_{t+1}$ to enable preliminary temporal modeling. In the second stage, to balance the performance of both present and future CSI prediction tasks, the loss function is redefined as:
\begin{equation}
\label{deqn_ex1a14}
{\mathcal L} = \lambda_1 \times MSE(\mathbf{H}_{t},\hat {\mathbf{H}}_{t}) + \lambda_2 \times MSE(\mathbf{H}_{t+1},\hat{\mathbf{H}}_{t+1}),
\end{equation}
where $\lambda_1$ and $\lambda_2$ are tunable hyperparameters. A smaller learning rate is used to fine-tune the LSTM and dual-input CSI reconstruction subnetwork, ensuring that future prediction capability is enhanced without degrading present prediction accuracy.

\section{Simulation Setup and Results}
This section evaluates the performance of the proposed ECB-P2WCP. The RT dataset, experimental configurations, and evaluation metrics are first introduced. Subsequently, the performance of ECB-P2WCP is presented and compared with other baseline methods in the context of present CSI prediction, followed by evaluations under more complex multi-user interference and localization error scenarios.

\begin{figure}[t]
\centering
\includegraphics[width=0.35\textwidth]{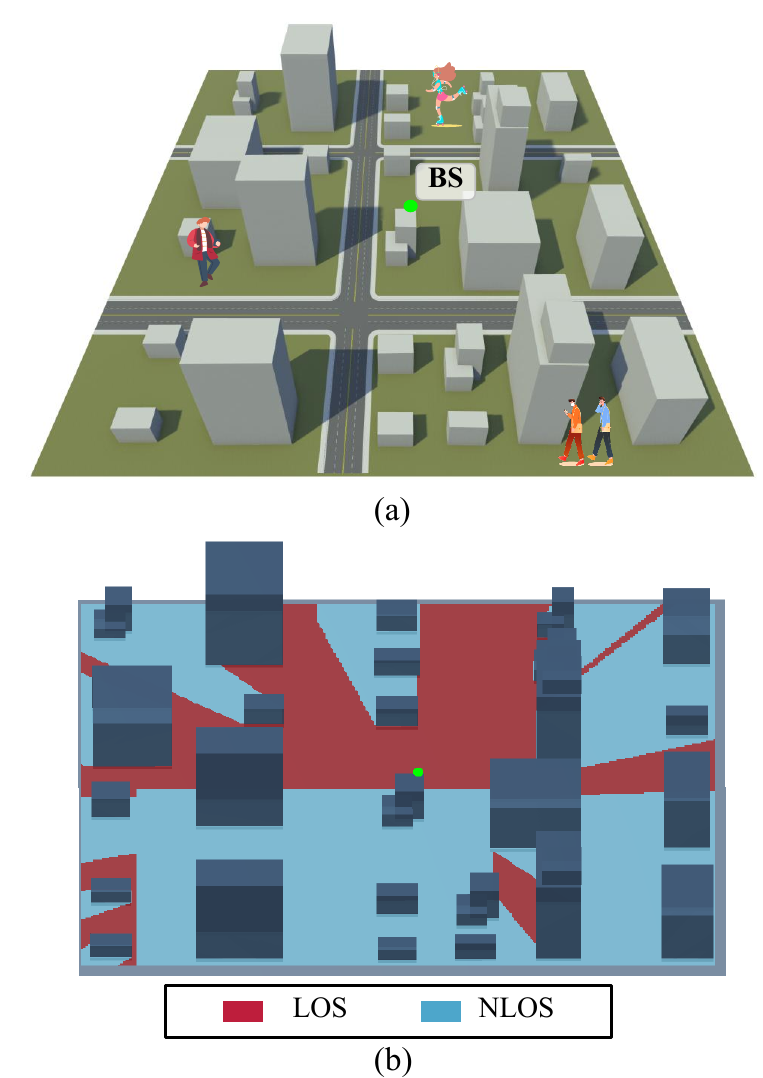}
\vspace{-1em}
\caption{RT simulation scenario. a) Environment layout; b) LoS/NLoS distribution. }
\label{rt}
\end{figure}

\subsection{Experiment Setup}
\textit{1) RT Scenario Setup:} 
As illustrated in Fig.~\ref{rt}(a), an outdoor UMa simulation environment is created in Blender, with 22 buildings of varying sizes distributed over a 240 m $\times$ 240 m area. The scene is subsequently imported into Wireless InSite~\cite{wi} to generate the CSI and ECB datasets. It is divided into 1,000 grids each of size 5 m $\times$ 5 m, and users are deployed according to the layout shown in Fig.~\ref{weks}. The data generation process follows the procedure described in Section~\ref{eb}, and the key simulation parameters are summarized in Table~\ref{RTparm}.

As shown in Fig.~\ref{rt}(b), the RT simulation covers 26,254 user locations, with approximately 40\% under LoS conditions. To enhance realism, CSCG noise is added to the generated channels, simulating two signal-to-noise ratio (SNR) scenarios: 10~dB (favorable) and 0~dB (challenging). In both cases, noise is applied to the channel samples at grid vertices and user positions, yielding noisy ECB and user CSI. To evaluate the future CSI prediction task, Doppler shifts are incorporated to emulate temporal channel variations. The number of previous time instances is set to $T=4$, with user mobility at 3 km/h and a time interval of 10 ms.

\begin{table}[t]
\caption{Channel and system simulation parameters.}
\label{RTparm}
\renewcommand{\arraystretch}{1.5}
\centering
\begin{tabular}{|>{\bfseries}l|l|}
\hline
\textbf{Parameters} & \textbf{Value} \\
\hline
Carrier frequency & 6.5 GHz \\
\hline
Bandwidth & 25 MHz \\
\hline
Subcarrier & 120 kHz, $N_{sc}$=208 \\
\hline
Number of antennas &  16 $\times$ 8 \\
\hline
Antenna interval & 0.5  wave-length \\
\hline
BS antenna& Istopic, vertical polarization. \\
\hline
User antenna& Istopic, vertical polarization. \\
\hline
BS location  & In center, height: 20 m\\
\hline
User location  & Interval: 1 m, height: 2 m \\
\hline
RT reflection order & 6 \\
\hline
RT diffraction order & 1 \\
\hline
RT propagation model & X3D \\
\hline
Maximum number of paths  & 25 \\

\hline
\end{tabular}
\end{table}

\begin{table}[t]
\caption{ECB-P2WNet parameters}
\label{DLparm}
\renewcommand{\arraystretch}{1.5}
\centering
\begin{tabular}{|>{\bfseries}l|l|}
\hline
\textbf{Parameters} & \textbf{Value} \\
\hline
Batch size & 32\\
\hline
Optimizer & Adam \\
\hline
Number of training data & 17500 \\
\hline
Number of validating data & 2500 \\
\hline
Number of testing data & 5000 \\
\hline
Activation function & LeakyReLU, coefficien=0.0005 \\
\hline
\end{tabular}
\end{table}

\textit{2) Method Parameters Setup and Performance Evaluation Metrics:} To evaluate the effectiveness of the proposed ECB-P2WCP, its performance is compared with the following four baseline methods. The linear minimum mean square error (LMMSE) estimator assumes prior knowledge of the second-order statistics of the entire scenario. It reconstructs the CSI as: $\mathbf{\hat h}_{\text{flat}} = {R_{h{h^0}}}R_{{h^0}{h^0}}^{ - 1}\mathbf{h}^0_{\text{flat}}$, where ${R_{h{h^0}}}=\frac{1}{N}\sum_{i=1}^{N} {{\mathbf h_i}\mathbf h{{_i^0}^T}} $ and ${R_{{h^0}{h^0}}}=\frac{1}{N}\sum_{i=1}^{N} {\mathbf h_i^0 \mathbf h{{_i^0}^T}}  + {\sigma ^2}\mathbf I$ represents the cross-correlation and auto-correlation matrices computed from the vertices channels. Here, $N$ denotes the number of known channels, and $\sigma ^2$ is the noise power. Besides, $\mathbf{I}$ denotes the unit matrix. 

The ideal ECB-based projection reconstruction method (IECB-PR) assumes an ideal ECB extracted from noiseless vertex channels. It utilizes the algorithm in Eq.~\eqref{e2} and Eq.~\eqref{e3} to acquire $\mathbf{\hat H}$, serving as the performance upper bound for PR methods. In contrast, the ECB-based projection reconstruction (ECB-PR) method incorporates realistic noise in both ECB and CSI, providing a more practical reference. Additionally, P2WCP is included as a representative conventional AI-based CSI prediction baseline. It excludes the ECB feature extraction subnetwork and the concatenation operation in the dual-input CSI reconstruction subnetwork, resulting in a model that relies solely on the partial channel matrix as input.

The dataset is divided into training, validation, and test sets in a ratio of 7:1:2, with data from 1,000 grids selected sequentially and randomly shuffled. The learning rate for the ECB feature extraction subnetwork is set to 0.0006, and for both the initial reconstruction subnetwork and the dual-input CSI reconstruction subnetwork, it is set to 0.0004. The model is trained for 200 epochs. For future CSI prediction, in the first training stage, the learning rate of the LSTM is set to 0.0003 and the model is trained for 100 epochs. In the second stage, the loss function is weighted with $\lambda_1=0.7$ and $\lambda_2=0.3$, and both the LSTM and the dual-input subnetwork are fine-tuned with a learning rate of 0.00005 for 50 epochs. Additional parameter settings are summarized in Table~\ref{DLparm}.

\begin{table*}[t]
\centering
\caption{Simulation Results for present CSI prediction.}
\label{p1}
\renewcommand{\arraystretch}{1.5}
\setlength\cellspacetoplimit{4pt}
\setlength\cellspacebottomlimit{4pt}
\begin{tabular}{|>{\centering\arraybackslash}p{2cm}|>{\centering\arraybackslash}p{2cm}|>{\centering\arraybackslash}p{1.5cm}|>{\centering\arraybackslash}p{1.5cm}|>{\centering\arraybackslash}p{1.5cm}|>{\centering\arraybackslash}p{1.5cm}|>{\centering\arraybackslash}p{1.5cm}|>{\centering\arraybackslash}p{1.5cm}|}
\hline
\multirow{2}{*}{\textbf{Pilot Ratio}} &\multirow{2}{*}{\textbf{Method}} & \multicolumn{3}{c|}{\textbf{10 dB}} & \multicolumn{3}{c|}{\textbf{0 dB}} \\
\cline{3-8}
 & &\textbf{NMSE} &\textbf{\makecell[c]{CS}} & \textbf{\makecell[c]{AAR \\ (bit/s/Hz)}} & \textbf{NMSE} &\textbf{\makecell[c]{CS}} & \textbf{\makecell[c]{AAR \\ (bit/s/Hz)}} \\
\hline
\multirow{5}{*}{1/4} &LMMSE  & 0.0710 & 0.9645 & 3.3608  & 0.5563 & 0.7655 & 0.6652\\
\cline{2-8}
& IECB-PR  &0.0278 & 0.9870 & 3.4283 & 0.0300 & 0.9861 & 0.9822\\
\cline{2-8}
 & ECB-PR&0.0720 & 0.9664 & 3.3720 & 0.2084 & 0.8993 & 0.8607\\
 \cline{2-8}
 & P2WCP  &0.0058 & 0.9972 & 3.452 & 0.0228 & 0.9889 & 0.9841\\
\cline{2-8}
 & ECB-P2WCP&0.0047 & 0.9977 & 3.453 & 0.0173 & 0.9916 & 0.9876\\
\hline
\multirow{5}{*}{1/8} & LMMSE&0.0999 & 0.9494 & 3.3192 & 0.7099 & 0.7101 & 0.5936\\
\cline{2-8}
& IECB-PR  &0.0320 & 0.9853 & 3.4265 & 0.0366 & 0.9834 & 0.9800\\
\cline{2-8}
 & ECB-PR &0.0766 & 0.9646 & 3.3690 & 0.2144 & 0.8966 & 0.8571\\
 \cline{2-8}
 & P2WCP &0.0106 & 0.9951 & 3.447 & 0.0394 & 0.981 & 0.9727 \\
\cline{2-8}
 & ECB-P2WCP &0.0068 & 0.9967 & 3.451 & 0.0267 & 0.9871 & 0.9818\\
\hline
\multirow{5}{*}{1/16} & LMMSE &0.1663 & 0.9115 & 3.2223	 & 0.7100 & 0.6678 & 0.5580\\
\cline{2-8}
& IECB-PR &0.0673 & 0.9696 & 3.3995 & 0.0764 & 0.9660 & 0.9641\\
\cline{2-8}
 & ECB-PR &0.1035 & 0.9526 & 3.3456 & 0.2349 & 0.8867 & 0.8453\\
 \cline{2-8}
 & P2WCP &0.0264 & 0.987 & 3.426 & 0.0836 & 0.9576 & 0.9407\\
\cline{2-8}
 & ECB-P2WCP &0.0162 & 0.992 & 3.439 & 0.0476 & 0.9768 & 0.9679\\
\hline
\multirow{5}{*}{1/32} & LMMSE &0.3079 & 0.8249 & 3.0266 & 0.7567 & 0.5875 & 	0.4899\\
\cline{2-8}
 & IECB-PR &0.1242 & 0.9463 & 3.3459 & 0.1390 & 0.9399 & 0.9339\\
\cline{2-8}
 & ECB-PR &0.1375 & 0.9370 & 3.3033 & 0.2599 & 0.8735 & 0.8270 \\
 \cline{2-8}
  & P2WCP &0.0749 & 0.9622 & 3.36 & 0.1834 & 0.902 & 0.8728\\
\cline{2-8}
 & ECB-P2WCP &\textbf{0.0357} & \textbf{0.9823} & \textbf{3.414} & \textbf{0.0823} & \textbf{0.9583} & \textbf{0.9441}\\
\hline
\end{tabular}
\end{table*}

\begin{figure*}[t]
\centering
\includegraphics[width=0.95\textwidth]{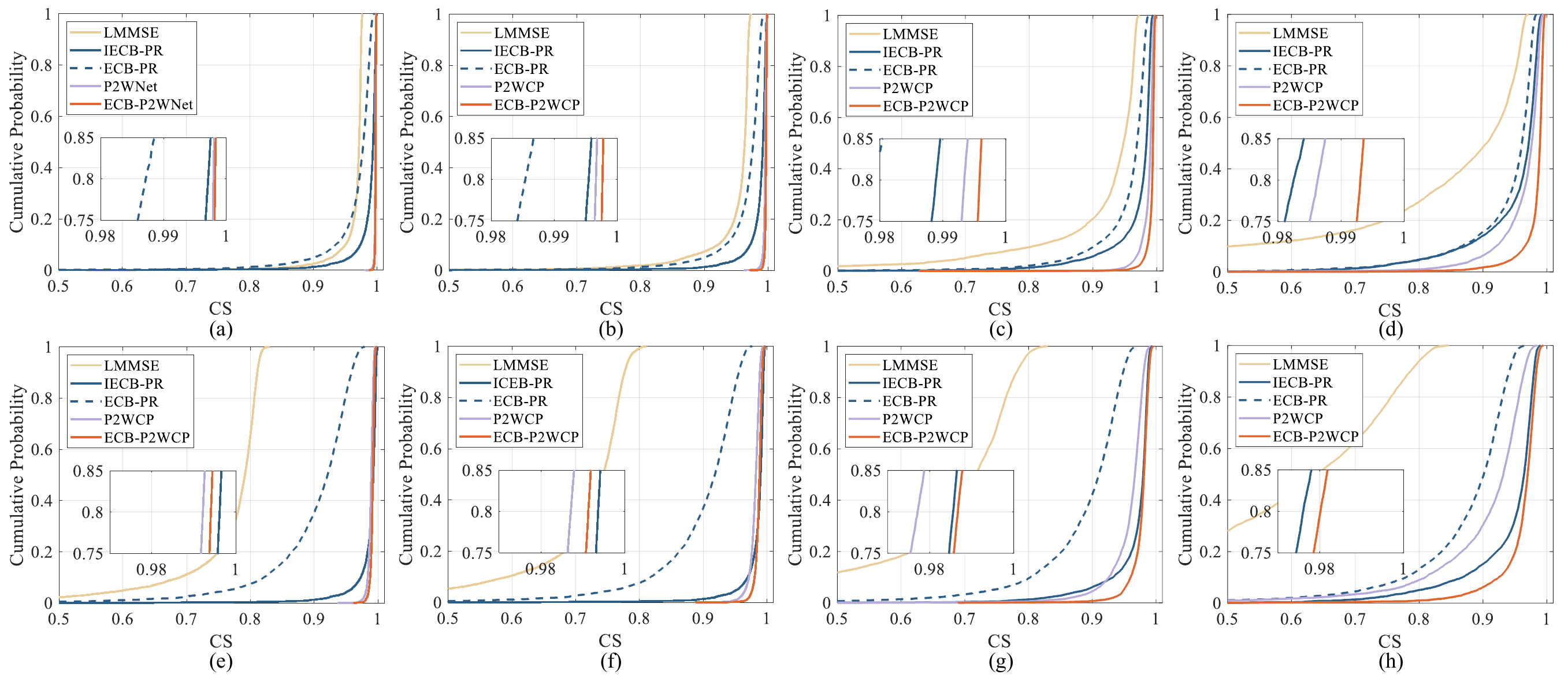}
\vspace{-1em}
\caption{CDF of CS between the predicted CSI and the true value. a) 10 dB SNR, 1/4 pilot ratio; b) 10 dB SNR, 1/8 pilot ratio; c) 10 dB SNR, 1/16 pilot ratio; d) 10 dB SNR, 1/32 pilot ratio; e) 0 dB SNR, 1/4 pilot ratio; f) 0 dB SNR, 1/8 pilot ratio; g) 0 dB SNR, 1/16 pilot ratio; h) 0 dB SNR, 1/32 pilot ratio.}
\label{corr}
\end{figure*}

Next, the performance of all methods is comprehensively evaluated in terms of CS and normalized mean square error (NMSE), with NMSE defined as:
\vspace{-1em}
\begin{equation}
\label{deqn_ex1a1}
NMSE=\frac{\left\|\mathbf{H}-\widehat{\mathbf{H}}\right\|_{F}^{2}}{\left\|\mathbf{H}\right\|_{F}^{2}}.
\end{equation}
Under beamforming-based downlink transmission, the average achievable rate (AAR) is given by:

\vspace{-1em}
\begin{equation}
\label{eq:aar}
\text{AAR} = \frac{1}{N_{sc}} \sum_{k=1}^{N_{sc}} \log_2 \left( 1 + \frac{ | \hat{\mathbf{h}}_k^H \mathbf{h}_k |^2 }{ \| \hat{\mathbf{h}}_k \|_2^2 \| {\mathbf{h}}_k \|_2^2 } \cdot \text{SNR} \right),
\end{equation}
where ${\mathbf{h}}_k \in {\mathbb{C}^{{M}_{t}}}$ denotes the actual CSI of the $k$-th subcarrier, and $\mathbf{\hat h}_{k}$ is the predicted value.

\vspace{-1em}

\subsection{Results and Discussions}
\begin{figure*}[t]
\centering
\includegraphics[width=0.95\textwidth]{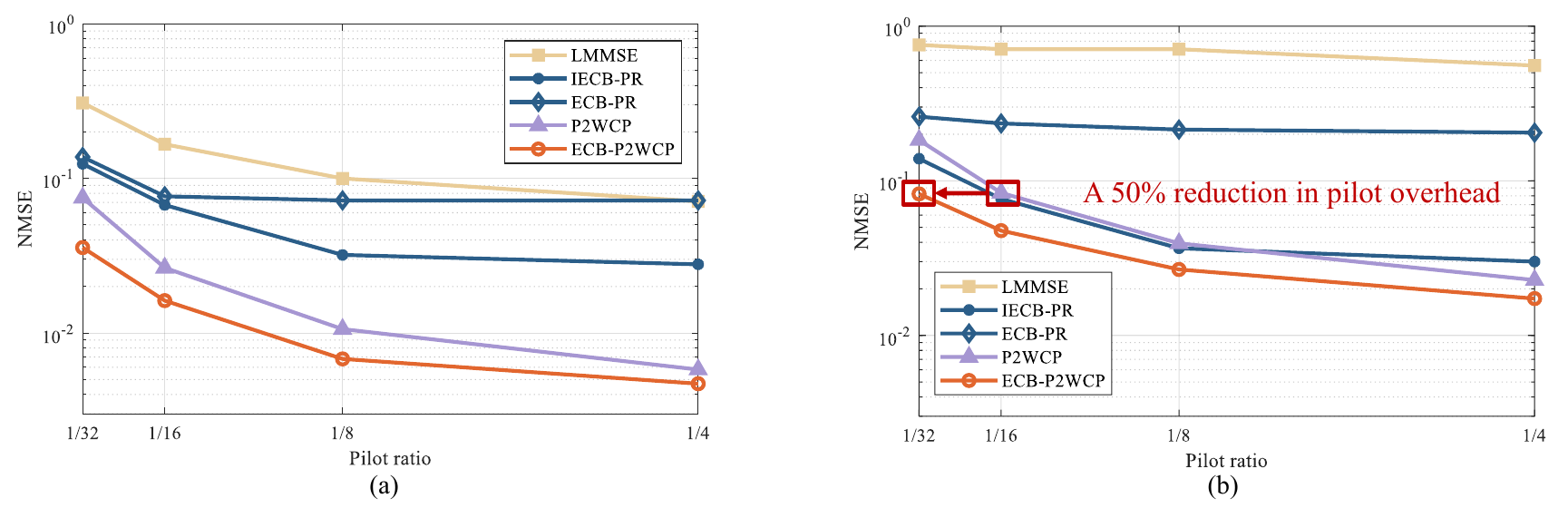}
\vspace{-1em}
\caption{NMSEs between the predicted CSI and the true value. a) 10 dB SNR; b) 0 dB SNR.}
\label{nmse}
\end{figure*}

\textit{1) Low-Overhead Present CSI Prediction:} 
Firstly, the performance of the proposed ECB-P2WCP in predicting present CSI is evaluated. Following the minimum frequency-domain pilot ratio (1/4) specified in 3GPP TR 38.211~\cite{38211}, lower pilot ratios of 1/4, 1/8, 1/16, and 1/32 are progressively examined, corresponding to $N_p$ = 52, 26, 13, and 7, respectively. As discussed in Section~\ref{eb}, the basis number $N_{b}=15$ is selected to ensure high prediction accuracy. Under this setup, the total inference time of the three subnetworks is 2.4 ms. If the $\mathbf{F}_b$ are precomputed and stored at the BS, the latency is reduced to 1.2 ms, demonstrating the feasibility of online deployment with 15 basis vectors.

Table~\ref{p1} summarizes the performance of different methods. The impact of SNR on prediction performance is first examined. Under a 10 dB SNR, all methods achieve high mean CS values ($>0.9$), with DL-based approaches (P2WCP and ECB-P2WCP) consistently exceeding 0.95 by effectively capturing internal correlations within the CSI. However, when the SNR drops to 0 dB, noise becomes more prominent and significantly degrades the performance of traditional methods. In particular, LMMSE experiences a notable decline, with CS falling below 0.6 at low pilot ratios. Interestingly, IECB-PR maintains competitive performance when the channel subspace basis is accurately extracted, occasionally even outperforming P2WCP and ECB-P2WCP. In contrast, ECB-PR proves less robust under noisy conditions when the ECB contains inaccuracies. Under these more realistic conditions, ECB-P2WCP outperforms ECB-PR with a 9\% improvement in CS and a 10\%–12\% gain in AAR, demonstrating the advantage of DL-based approaches over conventional mathematical methods.

\begin{figure}[t]
\centering
\includegraphics[width=0.4\textwidth]{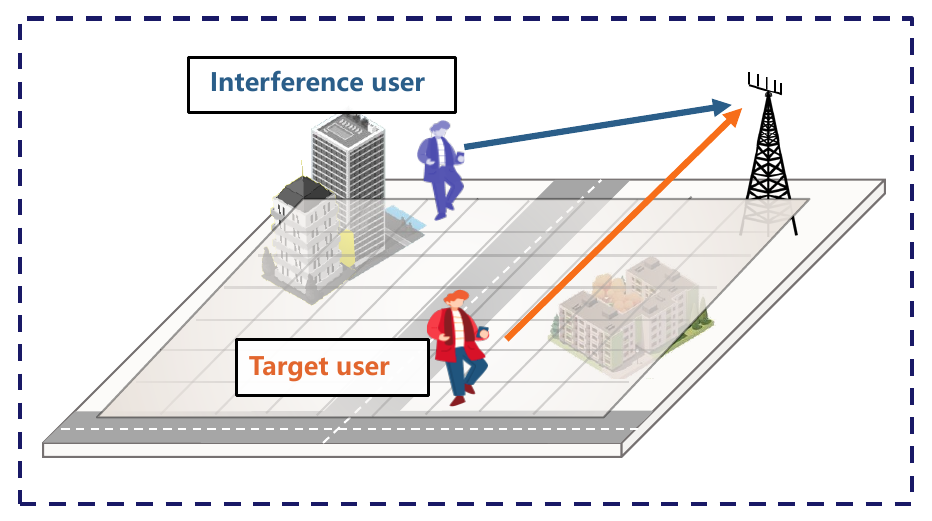}
\vspace{-0.5em}
\caption{The construction of a multi-user interference scenario}
\label{inter}
\end{figure}

The impact of pilot ratios is then evaluated to highlight the advantages of ECB-P2WCP further. Fig.~\ref{corr} presents the CS CDFs under various pilot ratios. A clear trend is observed: ECB-P2WCP gradually outperforms IECB-PR as the pilot ratio decreases. Under a 10 dB SNR, IECB-PR retains an advantage only at the $1/4$ pilot ratio. At 0 dB SNR, ECB-P2WCP slightly underperforms IECB-PR at $1/4$ ratio. However, at $1/8$, while the two methods perform similarly at the 50\% CDF point, ECB-P2WCP provides better overall average performance. At lower pilot ratios ($1/16$ and $1/32$), ECB-P2WCP surpasses IECB-PR, with the performance gap widening as the pilot ratio decreases. These results confirm that ECB-P2WCP excels when pilot information is limited. Its DL-based architecture effectively captures complex spatial–frequency structures from sparse inputs and remains robust to ECB inaccuracies through adaptive feature learning. 

Finally, the potential of environment information to reduce pilot overhead is investigated. As shown in Fig.~\ref{nmse}, when the pilot ratio is extremely low, P2WCP suffers significant performance degradation due to limited observations, as it relies solely on data-driven learning. In contrast, ECB-P2WCP leverages environmental priors to mitigate this issue, and its performance gain increases as the pilot ratio decreases. At a 1/32 pilot ratio and 10 dB SNR, ECB-P2WCP achieves NMSE improvements of 9.36, 5.41, 5.86, and 3.22 dB over the four baselines. At 0 dB SNR, the gains are 9.64, 2.28, 4.99, and 3.48 dB, respectively. These results indicate that ECB-P2WCP maintains strong performance with fewer pilots. Comparing with P2WCP, it can reduce pilot overhead by 25\%–30\% at 10 dB and 40\%–50\% at 0 dB SNR to achieve the same NMSE.

Overall, the proposed ECB-P2WCP achieves the lowest NMSE and the highest CS and AAR among all evaluated methods, exhibiting strong robustness to both ECB and CSI noise. Even under the challenging condition of 0 dB SNR and $1/32$ pilot ratio, ECB-P2WCP retains nearly 95\% of the AAR upper bound, highlighting its practical applicability.

\textit{2) Robustness to Multi-User Interference:} 
From the previous results, IECB-PR maintains consistently high performance under both 10 dB and 0 dB SNR, indicating that CSCG noise can be effectively filtered out via projection. To further evaluate the robustness and applicability of ECB-P2WCP under more complex conditions, colored noise is introduced into the CSI data. Specifically, as illustrated in Fig.~\ref{inter}, a multi-user interference scenario is simulated by mixing CSI matrices from two nearby users with a weight $a$:
\begin{equation}
\label{deqn_ex1a2}
{\mathbf{H}}_{x-\text{interference}} = {\mathbf{H}}_{{x-\text{noise}}} + {{a}} \times {\mathbf{H}}_{{y-\text{noise}}},
\end{equation}
where $\mathbf{H}_{x-\text{noise}}$ and $\mathbf{H}_{y-\text{noise}}$ are noisy CSI from the target and interfering user, respectively.

\begin{figure*}[t]
\centering
\includegraphics[width=0.9\textwidth]{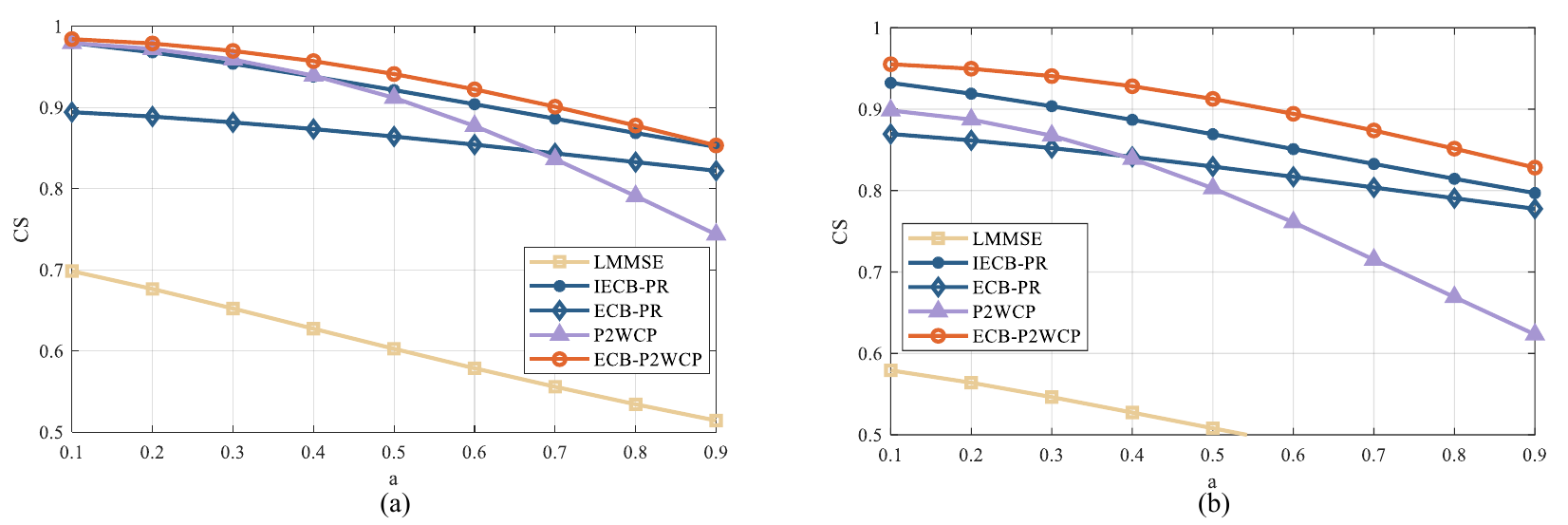}
\vspace{-1em}
\caption{CSs at multi-user interference scenario. a) pilot ratio of 1/8; b) pilot ratio of 1/32.}
\label{inter-corr}
\end{figure*}

Two representative cases from the 0 dB results in Table~\ref{p1} are selected: one with relatively limited performance gain for ECB-P2WCP (pilot ratio of 1/8), and one with more significant gain (pilot ratio of 1/32). These cases are tested using interference-free pretrained models. As shown in Fig.~\ref{inter-corr}, ECB-P2WCP consistently outperforms all baselines as $a$ increases from 0.1 to 0.9, demonstrating strong robustness to interference. When $a \leq 0.5$, performance degradation is minimal, and the performance gap with IECB-PR gradually increases, reaching a CS gain of 2\%–5\%. As $a > 0.5$, interference increasingly disrupts CSI correlations, degrading all DL-based methods. Among them, P2WCP is more vulnerable due to its lack of integration with environment information. When $a = 0.9$, ECB-P2WCP achieves 1.9 dB and 2.7 dB reductions in NMSE, 11\% and 20\% improvements in CS, and 15\% and 22\% gains in AAR compared to P2WCP.

Overall, the IECB-PR and ECB-PR algorithms, which benefit from fixed physical model assumptions, maintain relatively stable performance by projecting components of the signal onto the correct subspace. However, ECB-P2WCP demonstrates superior robustness and accuracy under mild and moderate levels of multi-user interference. Even in high-interference conditions, the environment prior provided by the ECB enables effective compensation for the distortion caused by interference, offering higher practical value for deployment in complex wireless environment.

\textit{3) Robustness to GPS Localization Errors:} Beyond CSI interference, the impact of errors in ECB selection is also evaluated. In the practical deployment of ECB-P2WCP, the user’s grid index is determined based on dynamic GPS coordinates, and the corresponding ECB is selected as the environmental prior. However, localization errors may cause mismatches between the selected ECB and the actual environment, resulting in reduced prediction accuracy.

\begin{figure*}[t]
\centering
\includegraphics[width=1\textwidth]{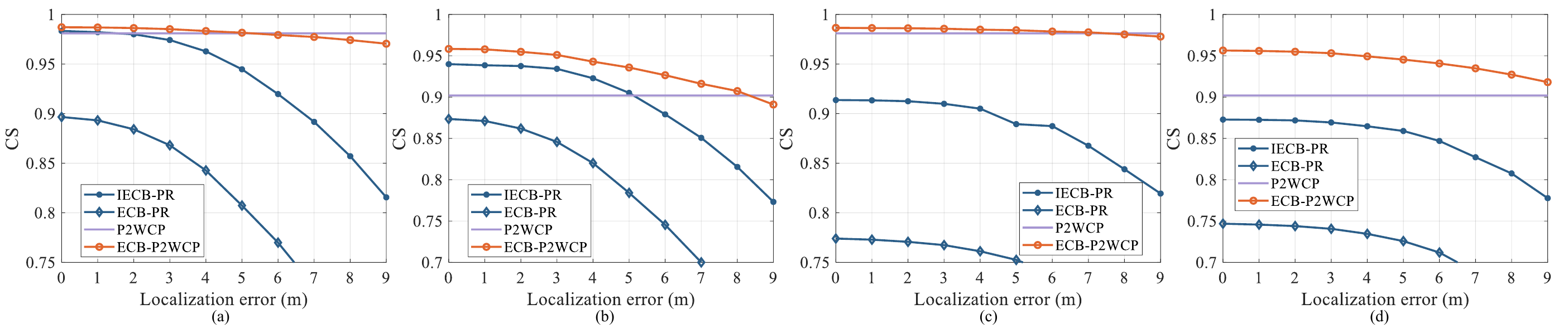}
\vspace{-2em}
\caption{CSs at the ECB selection error scenario. a) grid size of 5 m, pilot ratio of 1/8; b) grid size of 5 m, pilot ratio of 1/32; c) grid size of 10 m, pilot ratio of 1/8; d) grid size of 10 m, pilot ratio of 1/32.}
\label{locerr}
\end{figure*}

To assess this effect, random localization errors are introduced, with directions uniformly distributed and magnitudes ranging from 1 m to 10 m. The evaluation follows the same setup as in point 2), with the SNR set to 0 dB and pilot ratios of 1/8 and 1/32. Fig.~\ref{locerr} (a) and (b) present the performance of three ECB-based algorithms (IECB-PR, ECB-PR, and ECB-P2WCP) under different levels of GPS localization error. Since P2WCP does not rely on GPS-based ECB acquisition, its performance remains constant across all error levels. LMMSE also yields a constant CS but is excluded from the figure due to its low performance.

The results indicate that all ECB-based methods maintain stable performance under small localization errors ($\leq$ 2 m), with negligible degradation. As the error increases, IECB-PR and ECB-PR exhibit more rapid performance loss, whereas ECB-P2WCP remains relatively robust. For example, at a 3 m error with a 1/8 pilot ratio, ECB-P2WCP exhibits only a 0.56 dB increase in NMSE, a 0.19\% decrease in CS, and a 0.27\% reduction in AAR. When the localization error reaches 5 m, its performance approaches that of P2WCP, suggesting that the benefit of ECB is diminished. Similar trends are observed at a 1/32 pilot ratio: performance remains stable up to 2.5 m, while beyond 8 m, the ECB becomes ineffective in providing useful prior information.

\textit{4) Robustness to Spatial Granularity of ECB:}
Fine-grained spatial partitioning in large-scale outdoor scenarios can introduce considerable storage and computational overhead. To evaluate the robustness of ECB-P2WCP to ECB granularity, the impact of increasing the grid size used for ECB extraction is investigated. Specifically, the grid size is increased from 5 m to 10 m by merging four adjacent grids, and the ECB is re-extracted following the procedure described in Section~\ref{eb}. All three ECB-based algorithms are then re-evaluated using the updated basis.

Coarser grids improve spatial tolerance to localization errors, as users with moderate positioning deviations are more likely to remain within the correct grid. However, this comes at the cost of reduced ECB representativeness. As illustrated in Fig.~\ref{locerr} (c) and (d), IECB-PR and ECB-PR exhibit noticeable performance degradation under perfect localization conditions, with an average CS drop of approximately 7\%. Since these methods rely solely on basis fidelity, they are more vulnerable to reduced spatial resolution. While larger grids offer improved robustness to localization errors, they do not fully compensate for the accuracy loss, resulting in a decrease in overall performance.

In contrast to traditional methods, ECB-P2WCP exhibits minimal sensitivity to reductions in spatial resolution. Under perfect localization conditions, its performance after fine-tuning remains comparable to that under the 5 m grid. Compared to the results in Fig.~\ref{locerr} (a) and (b), the coarser 10 m grid yields improved robustness at the same error levels, with negligible performance loss when the error is below 3 m. At a pilot ratio of 1/8, the tolerable localization error before a 0.5 dB NMSE increase extends from 3 m to 3.5 m. The threshold at which the ECB no longer provides performance gain increases from 5 m to 8 m. At a 1/32 pilot ratio, the 0.5 dB NMSE increase threshold extends to 3 m, and even with a 10 m localization error, ECB-P2WCP continues to outperform P2WCP. These results demonstrate the effectiveness of leveraging environmental priors and the strong generalization capability of ECB-P2WCP.

Overall, ECB-P2WCP effectively adapts to coarser grid resolutions through fine-tuning, achieving increased localization tolerance while maintaining high prediction accuracy.

\begin{table}[t]
\centering
\renewcommand{\arraystretch}{1.5}
\caption{Simulation Results for Future CSI Prediction}
\label{p2}
\begin{tabular}{|>{\centering\arraybackslash}m{1cm}|
                >{\centering\arraybackslash}m{1.7cm}|
                >{\centering\arraybackslash}m{1.1cm}|
                >{\centering\arraybackslash}m{1.1cm}|
                >{\centering\arraybackslash}m{1.1cm}|}
\hline
\textbf{\makecell[c]{Setting}} & \textbf{Method} & \textbf{NMSE} & \textbf{CS} & \textbf{\makecell[c]{AAR \\ (bit/s/Hz)}}\\
\hline

\hline
\multirow{4}{*}{\makecell[c]{10 dB \\ 1/8}} & IECB-PR & 0.3282 & 0.9216 & 3.2503 \\
\cline{2-5}
 & ECB-PR & 0.3477 & 0.9058 & 3.2043 \\
 \cline{2-5}
 & P2WCP & 0.0341 & 0.9842 & 3.418 \\
 \cline{2-5}
 & ECB-P2WCP & 0.0232 & 0.9895 & 3.432 \\
\hline
\multirow{4}{*}{\makecell[c]{10 dB \\ 1/32}} & IECB-PR & 0.3803 & 0.8812 & 3.1642 \\
\cline{2-5}
 & ECB-PR & 0.3883 & 0.8756 & 3.1294 \\
 \cline{2-5}
 & P2WCP & 0.114 & 0.9431 & 3.3058 \\
 \cline{2-5}
 & ECB-P2WCP & 0.0617 & 0.9705 & 3.383 \\
\hline
\multirow{4}{*}{\makecell[c]{0 dB \\ 1/8}} & IECB-PR & 0.3282 & 0.9216 & 0.8960 \\
\cline{2-5}
 & ECB-PR & 0.4336 & 0.8523 & 0.8001 \\
 \cline{2-5}
 & P2WCP & 0.0795 & 0.963 & 0.9479 \\
 \cline{2-5}
 & ECB-P2WCP & 0.0565 & 0.9741 & 0.9637 \\
 \hline
\multirow{4}{*}{\makecell[c]{0 dB \\ 1/32}} & IECB-PR & 0.3803 & 0.8812 & 0.8566 \\
\cline{2-5}
 & ECB-PR & 0.4634 & 0.8303 & 0.7723 \\
 \cline{2-5}
 & P2WCP & 0.213 & 0.8928 & 0.857 \\
 \cline{2-5}
 & ECB-P2WCP & 0.121 &  \textbf{0.9431} &  \textbf{0.9225} \\
\hline
\end{tabular}
\end{table}

\textit{5) Future CSI Prediction Without Additional Pilots:}
To further reduce pilot overhead, the performance of ECB-P2WCP in future CSI prediction is evaluated. In this setting, P2WCP and ECB-P2WCP are used to predict the CSI at the fourth time instance based on the previous three $\mathbf{H^0}$. As a baseline, IECB-PR and ECB-PR reuse the third-step CSI to approximate the next instant, illustrating the impact of channel aging.

As shown in Table~\ref{p2}, DL-based methods significantly outperform the traditional baselines. ECB-P2WCP achieves up to 10\% higher CS than ECB-PR and exceeds IECB-PR by 6\%. Compared to P2WCP, it yields NMSE gains of 1.5 dB and 2.5 dB, along with 1–4.5\% improvements in both CS and AAR at pilot ratios of 1/8 and 1/32. Although the predicted $\mathbf{H}_\text{ini}$ using LSTM is less accurate than in the present CSI prediction case, ECB-P2WCP still maintains an average CS above 0.97 under favorable conditions and completes prediction within 1.3 ms. Even under the most challenging scenario (0 dB SNR, 1/32 pilot ratio), it achieves a CS close to 0.95 and over 90\% of the AAR upper bound, demonstrating strong prediction capability.

These results confirm that ECB-P2WCP can effectively predict the CSI of the next channel coherence time, enabling a further 25\% reduction in pilot overhead while maintaining acceptable prediction accuracy. This offers a practical approach to reducing resource consumption, mitigating channel aging, and supporting proactive communication strategies.

\section{Conclusion}
In this paper, we propose the ECB-P2WCP, a novel method that integrates dynamic and static information to enable low-overhead CSI prediction. By analyzing channel characteristics across different user locations, we develop an environment-specific subspace extraction method that captures most of the channel variations without requiring real-time updates during deployment. Then, a CNN- and transformer-based neural network is designed to perform ECB-aided CSI prediction. During practical deployment, users only need to transmit their GPS coordinates along with pilot signals on a small portion of the frequency domain channel. 

Simulation results show that the proposed ECB offers significant advantages under low SNR and sparse pilot conditions, reducing pilot overhead by up to 50\%. The prior knowledge encoded in the ECB also compensates for performance degradation caused by multi-user interference. Furthermore, the model tolerates localization errors of up to 3 meters, making it well-suited for outdoor scenarios. In the future CSI prediction task, the proposed method is able to forecast the next CSI value within 1.3 ms, allowing the system to mitigate the effects of channel aging.

Looking forward, the ECB-based environment representation can be extended to time-varying or mobility-rich scenarios through adaptive basis updates. As wireless communication systems evolve toward intelligent and autonomous control, the proposed ECB-P2WCP offers a practical approach for bridging physical environments and data-driven CSI prediction, enabling more proactive and environment-intelligent transmission strategies.

\vspace{-1.5em}
\bibliographystyle{IEEEtran}
\bibliography{ref}

\begin{thebibliography}{10}
\providecommand{\url}[1]{#1}
\csname url@samestyle\endcsname
\providecommand{\newblock}{\relax}
\providecommand{\bibinfo}[2]{#2}
\providecommand{\BIBentrySTDinterwordspacing}{\spaceskip=0pt\relax}
\providecommand{\BIBentryALTinterwordstretchfactor}{4}
\providecommand{\BIBentryALTinterwordspacing}{\spaceskip=\fontdimen2\font plus
\BIBentryALTinterwordstretchfactor\fontdimen3\font minus \fontdimen4\font\relax}
\providecommand{\BIBforeignlanguage}[2]{{%
\expandafter\ifx\csname l@#1\endcsname\relax
\typeout{** WARNING: IEEEtran.bst: No hyphenation pattern has been}%
\typeout{** loaded for the language `#1'. Using the pattern for}%
\typeout{** the default language instead.}%
\else
\language=\csname l@#1\endcsname
\fi
#2}}
\providecommand{\BIBdecl}{\relax}
\BIBdecl

\bibitem{lgy6G}
G.~Liu, Y.~Huang, N.~Li \emph{et~al.}, ``{Vision, requirements and network architecture of 6G mobile network beyond 2030},'' \emph{China Commun.}, vol.~17, no.~9, pp. 92--104, Sep. 2020.

\bibitem{6G}
J.~Zhang, J.~Lin, P.~Tang \emph{et~al.}, ``{Channel measurement, modeling, and simulation for 6G: A survey and tutorial},'' \emph{arXiv preprint arXiv:2305.16616}, May 2023.

\bibitem{lhq}
H.~Lu, Y.~Zeng, C.~You \emph{et~al.}, ``{A tutorial on near-field XL-MIMO communications toward 6G},'' \emph{IEEE Commun. Surv. Tutor.}, vol.~26, no.~4, pp. 2213--2257, Apr. 2024.

\bibitem{est1}
F.~Tufvesson and T.~Maseng, ``{Pilot assisted channel estimation for OFDM in mobile cellular systems},'' \emph{Proc. IEEE Veh. Technol. Conf. (VTC)}, vol.~3, pp. 1639--1643 vol.3, May 1997.

\bibitem{ZZ_AI}
Z.~Zhang, J.~Zhang, Y.~Zhang \emph{et~al.}, ``{AI}-based time\nobreakdash-, frequency\nobreakdash-, and space-domain channel extrapolation for {6G: Opportunities} and challenges,'' \emph{IEEE Veh. Technol. Mag.}, vol.~18, no.~1, pp. 29--39, Mar. 2023.

\bibitem{zz_drl}
Z.~\vspace{0mm}Zhang, J.~Zhang, Y.~Zhang \emph{et~al.}, ``{Deep reinforcement learning based dynamic beam selection in dual-band communication systems},'' \emph{IEEE Trans. Wireless Commun.}, vol.~23, no.~4, pp. 2591--2606, Apr. 2024.

\bibitem{chazhi}
S.~Coleri, M.~Ergen, A.~Puri, and A.~Bahai, ``{Channel estimation techniques based on pilot arrangement in OFDM systems},'' \emph{IEEE Trans. Broadcast.}, vol.~48, no.~3, pp. 223--229, Sep. 2002.

\bibitem{14}
Y.~Liao, Y.~Hua, and Y.~Cai, ``{Deep learning based channel estimation algorithm for fast time-varying MIMO-OFDM systems},'' \emph{IEEE Commun. Lett.}, vol.~24, no.~3, pp. 572--576, Mar. 2020.

\bibitem{czr_jsac}
Z.~Chen, Z.~Zhang, Z.~Yang \emph{et~al.}, ``{Channel deduction: A new learning framework to acquire channel from outdated samples and coarse estimate},'' \emph{IEEE J. Sel. Areas Commun.}, vol.~43, no.~3, pp. 944--958, Mar. 2025.

\bibitem{twc_est}
M.~del Rosario and Z.~Ding, ``{Learning-based MIMO channel estimation under practical pilot pparsity and feedback compression},'' \emph{IEEE Trans. Wireless Commun.}, vol.~22, no.~2, pp. 1161--1174, Feb. 2023.

\bibitem{perdict}
G.~Nie, J.~Zhang, Y.~Zhang \emph{et~al.}, ``{A predictive 6G network with environment sensing enhancement: From radio wave propagation perspective},'' \emph{China Commun.}, vol.~19, no.~6, pp. 105--122, June 2022.

\bibitem{jianhua_eic}
J.~Zhang, L.~Yu, S.~Liu \emph{et~al.}, ``Wireless environmental information theory: A new paradigm towards {6G} online and proactive environment intelligence communication,'' \emph{Engineering}, 2025.

\bibitem{dtc}
H.~Wang, J.~Zhang, G.~Nie \emph{et~al.}, ``{Digital twin channel for 6G: Concepts, architectures and potential applications},'' \emph{IEEE Commun. Mag.}, vol.~63, no.~3, pp. 24--30, Mar. 2025.

\bibitem{ckm2}
Y.~Zeng and X.~Xu, ``{Toward environment-aware 6G communications via channel knowledge map},'' \emph{IEEE Wirel. Commun.}, vol.~28, no.~3, pp. 84--91, Jun. 2021.

\bibitem{ckm_CSI}
X.~Wang, Y.~Shi, T.~Wang \emph{et~al.}, ``{Can channel knowledge map help to predict instantaneous MIMO channel state information?}'' in \emph{Proc. IEEE Wireless Commun. Netw. Conf. (WCNC)}, Apr. 2024, pp. 1--6.

\bibitem{gff_camera}
M.~Ouyang, F.~Gao, Y.~Wang \emph{et~al.}, ``{Computer vision-aided reconfigurable intelligent surface-based beam tracking: Prototyping and experimental results},'' \emph{IEEE Trans. Wireless Commun.}, vol.~22, no.~12, pp. 8681--8693, Dec. 2023.

\bibitem{lidar}
B.~Salehi, G.~Reus-Muns, D.~Roy \emph{et~al.}, ``{Deep learning on multimodal sensor data at the wireless edge for vehicular network},'' \emph{IEEE Trans. Veh. Technol.}, vol.~71, no.~7, pp. 7639--7655, July 2022.

\bibitem{mmff}
H.~Zhang, S.~Gao, X.~Cheng \emph{et~al.}, ``{Integrated sensing and communications toward proactive beamforming in mmWave V2I via multi-modal feature fusion (MMFF)},'' \emph{IEEE Trans. Wireless Commun.}, vol.~23, no.~11, pp. 15\,721--15\,735, Nov. 2024.

\bibitem{SYT_feature}
Y.~Sun, J.~Zhang, Y.~Zhang \emph{et~al.}, ``{Environment features-based model for path loss prediction},'' \emph{IEEE Wireless Commun. Lett.}, vol.~11, no.~9, pp. 2010--2014, Sep. 2022.

\bibitem{4steps}
J.~Zhang, Y.~Cai, L.~Yu \emph{et~al.}, ``{Four steps toward 6G AI-enabled air interface: Wireless environmental information sensing, feature, semantics, and knowledge},'' \emph{IEEE Commun. Mag.}, vol.~63, no.~8, pp. 56--62, August 2025.

\bibitem{REKP}
J.~Wang, J.~Zhang, Y.~Zhang \emph{et~al.}, ``{Radio environment knowledge pool for 6G digital twin channel},'' \emph{IEEE Commun. Mag.}, vol.~63, no.~5, pp. 158--164, May 2025.

\bibitem{zz5}
R.~W. Heath, N.~González-Prelcic, S.~Rangan \emph{et~al.}, ``{An overview of signal processing techniques for millimeter wave MIMO systems},'' \emph{IEEE J. Sel. Topics Signal Process.}, vol.~10, no.~3, pp. 436--453, Apr. 2016.

\bibitem{deepmimo}
A.~Alkhateeb, ``{DeepMIMO: A generic deep learning dataset for millimeter Wave and massive MIMO applications},'' \emph{arXiv preprint arXiv:1902.06435}, Feb. 2019.

\bibitem{scene-specific}
X.~Li, J.~Guo, C.-K. Wen \emph{et~al.}, ``{Auto-CsiNet: Scenario-customized automatic neural network architecture generation for massive MIMO CSI feedback},'' \emph{IEEE Trans. Wireless Commun.}, vol.~23, no.~10, pp. 14\,759--14\,775, Oct. 2024.

\bibitem{finger}
E.~Kupershtein, M.~Wax, and I.~Cohen, ``{Single-site emitter localization via multipath fingerprinting},'' \emph{IEEE Trans. Signal Process.}, vol.~61, no.~1, pp. 10--21, Jan. 2013.

\bibitem{gff_rank}
H.~Xie, F.~Gao, and S.~Jin, ``{An overview of low-rank channel estimation for massive MIMO systems},'' \emph{IEEE Access}, vol.~4, pp. 7313--7321, Nov. 2016.

\bibitem{zjj}
C.-K. Wen, W.-T. Shih, and S.~Jin, ``{Deep learning for massive MIMO CSI feedback},'' \emph{IEEE Wireless Commun. Lett.}, vol.~7, no.~5, pp. 748--751, Oct. 2018.

\bibitem{prior_csi}
B.~V. Boas, W.~Zirwas, and M.~Haardt, ``Machine learning for {CSI} recreation in the digital twin based on prior knowledge,'' \emph{IEEE Open J. Commun. Soc.}, vol.~3, pp. 1578--1590, Sep. 2022.

\bibitem{zy3}
D.~Wu, Y.~Zeng, S.~Jin, and R.~Zhang, ``{Environment-aware hybrid beamforming by leveraging channel knowledge map},'' \emph{IEEE Transactions on Wireless Communications}, vol.~23, no.~5, pp. 4990--5005, May 2024.

\bibitem{lqs}
A.~Ispas, C.~Schneider, G.~Ascheid \emph{et~al.}, ``{Analysis of the local quasi-stationarity of measured dual-polarized MIMO channels},'' \emph{IEEE Transactions on Vehicular Technology}, vol.~64, no.~8, pp. 3481--3493, Aug. 2015.

\bibitem{ckm_qyl}
Y.~Qiu, D.~Wu, and Y.~Zeng, ``{CKM-based environment-aware pilot reuse and channel estimation},'' in \emph{2024 16th International Conference on Wireless Communications and Signal Processing (WCSP)}, Oct. 2024, pp. 169--174.

\bibitem{pca}
H.~Hotelling, ``{Analysis of a complex of statistical variables into principal components},'' \emph{Journal of Educational Psychology}, vol.~24, no.~6, pp. 417--441, Sep. 1933.

\bibitem{lstm}
S.~Hochreiter and J.~Schmidhuber, ``{Long short-term memory},'' \emph{Neural Comput.}, vol.~9, no.~8, pp. 1735--1780, Nov. 1997.

\bibitem{cnn}
A.~Krizhevsky, I.~Sutskever, and G.~E. Hinton, ``{ImageNet classification with deep convolutional neural networks},'' in \emph{Advances in Neural Information Processing Systems}, vol.~25.\hskip 1em plus 0.5em minus 0.4em\relax Curran Associates, Inc., 2012.

\bibitem{transformer}
A.~Vaswani, N.~Shazeer, N.~Parmar \emph{et~al.}, ``{Attention is all you need},'' in \emph{Advances in Neural Information Processing Systems}, vol.~30.\hskip 1em plus 0.5em minus 0.4em\relax Curran Associates, Inc., Dec. 2017.

\bibitem{dps}
H.~Iris, B.~S. Veeling, and R.~{van Sloun}, ``\BIBforeignlanguage{English}{{Deep probabilistic subsampling for task-adaptive compressed sensing}},'' \emph{\BIBforeignlanguage{English}{Proc. Int. Conf. Learn. Represent. (ICLR)}}, Apr. 2020.

\bibitem{resnet}
K.~He, X.~Zhang, S.~Ren \emph{et~al.}, ``{Deep residual learning for image recognition},'' \emph{IEEE Trans. Pattern Anal. Mach. Intell.}, vol.~38, no.~12, pp. 770--778, Dec. 2016.

\bibitem{wi}
\BIBentryALTinterwordspacing
{Wireless InSite}, ``{Wireless inSite: radio propagation simulation software},'' 2023, accessed: 2025-02-27. [Online]. Available: \url{https://www.spectracomcorp.com/wireless-insite/}
\BIBentrySTDinterwordspacing

\bibitem{38211}
{3GPP}, ``{Physical channels and modulation},'' \emph{3GPP TR 38.211 Release 18}, Mar. 2024.

\end{thebibliography}


 



\vfill

\end{document}